\documentclass[twocol]{ametsocV6.1}

\usepackage{accents}

\usepackage{pdfpages} % include pdfs
\usepackage{pgffor} % for loops

\makeatletter
\AtBeginDocument{\let\LS@rot\@undefined}
\makeatother

\def\supplementfilename{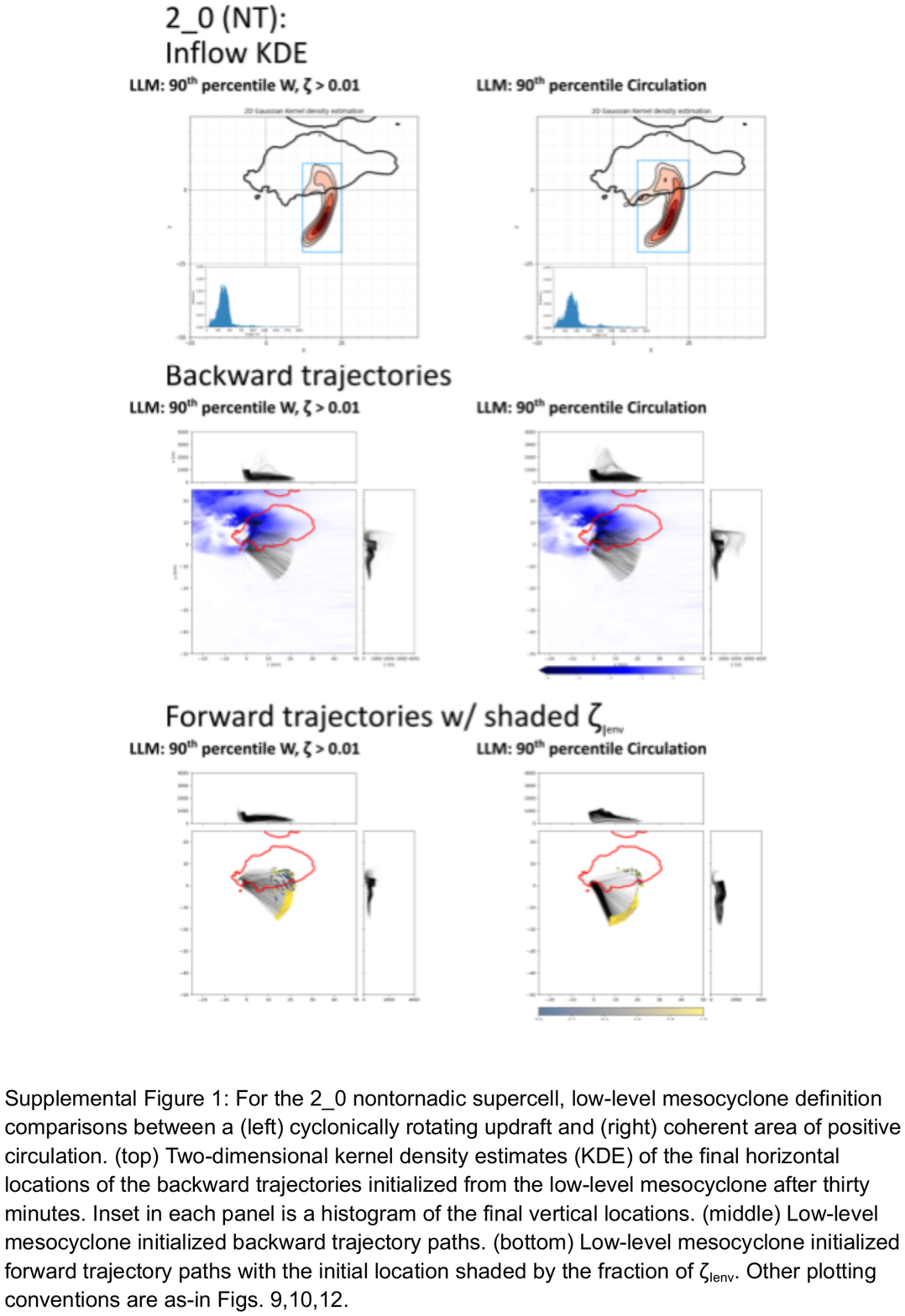}

\pdfximage{\supplementfilename}
\def\numbersupplementpages{\the\pdflastximagepages}

\newif\ifarXiv
\arXivtrue 
\newcommand{\ut}[1]{\underaccent{\tilde}{#1}}
\renewcommand{\vec}[1]{\ut{#1}}

\title{Supercell low-level mesocyclones: Origins of inflow and vorticity}

\authors{Brice E. Coffer,\aff{a}\correspondingauthor{Brice E. Coffer, becoffer@ncsu.edu}
Matthew D. Parker,\aff{a} 
John M. Peters,\aff{b} 
Andrew R. Wade\aff{c}
}
\affiliation{\aff{a}{Department of Marine, Earth, and Atmospheric Sciences, North Carolina State University, Raleigh, NC}\\
\aff{b}{Department of Meteorology and Atmospheric Science, The Pennsylvania State University, University Park, PA}\\
\aff{c}{Cooperative Institute for Severe and High-Impact Weather Research and Operations, University of Oklahoma, Norman, OK}\\
}

\abstract{The development of low-level mesocyclones in supercell thunderstorms has often been explained via the development of storm-generated streamwise vorticity along a baroclinic gradient in the forward flank of supercells. However, the ambient streamwise vorticity of the environment (often quantified via storm-relative helicity), especially near the ground, is particularly skillful at discriminating between nontornadic and tornadic supercells. This study investigates whether the origins of the inflow air into supercell low-level mesocyclones, both horizontally and vertically, can help explain the dynamical role of environmental versus storm-generated vorticity in the development of low-level mesocyclone rotation. Simulations of supercells, initialized with wind profiles common to supercell environments observed in nature, show that the air bound for the low-level mesocyclone primarily originates from the undisturbed, ambient environment, rather than from along the forward flank, and from very close to the ground, often in the lowest 200 - 400 m of the atmosphere. Given that the near-ground environmental air comprises the bulk of the inflow into low-level mesocyclones, this likely explains the forecast skill of environmental streamwise vorticity in the lowest few hundred meters of the atmosphere. The low-level mesocyclone does not appear to require much augmentation from the development of additional horizontal vorticity in the forward flank. Instead, the dominant contributor to vertical vorticity within the low-level mesocyclone is from the environmental horizontal vorticity. This study hopefully clarifies the development of low-level mesocyclones in supercells.}

\begin{document}

%% Necessary!
\maketitle

%%%%%%%%%%%%%%%%%%%%%%%%%%%%%%%%%%%%%%%%%%%%%%%%%%%%%%%%%%%%%%%%%%%%%
% SIGNIFICANCE STATEMENT/CAPSULE SUMMARY
%%%%%%%%%%%%%%%%%%%%%%%%%%%%%%%%%%%%%%%%%%%%%%%%%%%%%%%%%%%%%%%%%%%%%
%
% Significance Statement (all journals except BAMS)
%
\statement
Supercell thunderstorms produce the majority of tornadoes, and a defining characteristic of supercells is their rotating updraft, known as the ``mesocyclone’’. When the mesocyclone is stronger at lower altitudes, the likelihood of tornadoes increases. The purpose of this study is to understand if the rotation of the mesocyclone in supercells is due to horizontal spin present in the ambient environment or whether additional horizontal spin generated by the storm itself primarily drives this rotation. Our results suggest that inflow air into supercells, and low-level mesocyclone rotation, is mainly due to the properties of the environmental inflow air, especially near the ground. This hopefully will clarify how our community views the development of low-level mesocyclones in supercells. 

%%%%%%%%%%%%%%%%%%%%%%%%%%%%%%%%%%%%%%%%%%%%%%%%%%%%%%%%%%%%%%%%%%%%%
% MAIN BODY OF PAPER
%%%%%%%%%%%%%%%%%%%%%%%%%%%%%%%%%%%%%%%%%%%%%%%%%%%%%%%%%%%%%%%%%%%%%
%

\section{Introduction}

A defining characteristic of supercell thunderstorms is their mesocyclone, a quasi-steady region of vertical vorticity within the storm's updraft. This persistent feature contributes to the supercell's ability to produce a host of severe weather threats, including damaging nontornadic winds \citep{smith2012modes}, flash flooding \citep{nielsen2020observations}, large individual hail stones \citep{blair2017hail} and/or large accumulations of small hail \citep{kumjian2019small}, and tornadoes \citep{markowski2009tornadogenesis}.
%In addition to its physical importance, the mesocyclone is not only often visually spectacular, which in part, has led to the rise of storm chasing as a tourist attraction \citep{boulais2017severe}, but in Doppler radar data, the mesocyclone is also associated with a couplet of inbound and outbound radial velocities, a key part of identifying supercells in real-time nowcasting operations \citep{burgess1993tornado}. 
Conceptual models of supercells have all prominently featured the mesocyclone \citep{brandes1978mesocyclone,lemon1979severe,klemp1987dynamics,doswell1993tornadoes}, and temporal increases in mesocyclone rotation near the cloud-base have long been identified as a precursor of tornadoes \citep[e.g.,][]{brandes1993tornadic,burgess1993tornado,thompson2017damage}. %and is a fundamental component of storm spotter training programs \citep{moller1978storm}. 
Despite these connections between mesocyclones and tornadoes, the presence of a mesocyclone alone is a poor predictor of supercellular tornadogenesis \citep{trapp1999observations}. Although most tornadoes are associated with supercells, perhaps less than 15\% of mesocyclones are tornadic \citep{trapp2005reassessment, smith2012modes}. 

%\begin{figure*}[t]
%\centerline{\includegraphics[width=20pc]{Figure1.jpg}}
%\caption{A supercell near Imperial, Nebraska, on 27 May 2019 looking toward the northwest, labeled with relevant features of interest. (Photograph courtesy of S. Peake).} \label{conceptual}
%\end{figure*}

The process of supercellular tornadogenesis is often described as having three steps \citep{davies2015review}. First, an updraft needs to require rotation aloft (i.e., the development of a mesocyclone). %Although a singular step, the formation of the mesocyclone in the mid-tropsophere, compared to the lower troposphere, are regularly considered separate processes. 
It is well established that the updrafts of supercells initially acquire their rotation via the tilting of horizontal vorticity associated with the vertical shear of the environmental wind profile \citep[e.g.,][]{rotunno1982influence,davies1984streamwise,weisman2000shear}. Specifically, the tilting of streamwise horizontal vorticity (i.e., the component of vorticity aligned parallel to the motion of air in the storm-relative framework) is a necessary requirement for the updraft to acquire net positive vertical vorticity \citep{davies1984streamwise,davies2017roles,davies2022theory,dahl2017tilting}. %eAt lower altitudes, the development of the \emph{low-level} mesocyclone has largely been attributed to \emph{storm-generated streamwise horizontal vorticity associated with baroclinic gradients within the supercell itself}. The orientation of the forward-flank baroclinic zone and the storm-relative flow in this region of the storm is such that baroclinically generated vorticity is predominantly streamwise \citep{klemp1983study}, and as long as the temperature deficit within a storm’s forward flank outflow is not prohibitively excessive \citep{shabbott2006surface}, rain-cooled, streamwise vorticity-rich air is frequently ingested by the parent storm's main updraft \citep[e.g., the visual manifestation of the ``wall cloud'';][]{atkins2014observations}. 

The second step in tornadogenesis involves rotation developing at the ground. Surface vertical vorticity is thought to occur primarily through some combination of baroclinic \citep[e.g.,][]{davies1982observational,davies1993mesocyclogenesis,walko1993tornado,adlerman1999numerical,dahl2014imported,markowski2014influence, dahl2015near, parker2015production} and frictional \citep[e.g.,][]{schenkman2014tornadogenesis, markowski2016drag,mashiko2016numerical,roberts2016role,yokota2018important,fischer2022transition} generation of horizontal vorticity within downdrafts.
Following generation, this horizontal vorticity is subsequently tilted into the vertical very close to the surface \citep{rotunno2017near}, typically in cyclonically curved, descending air parcels embedded in the rear-flank outflow near the tip of the hook echo \citep{davies2022theory}. 

Both nontornadic and tornadic supercells readily acquire this sub-tornadic rotation near the surface \citep{parker2015production,coffer2017volatility}. Hence, steps 1 and 2 are necessary, but not sufficient, for tornadogenesis. The third and final step in tornadogenesis is the contraction and intensification of coherent areas of large circulation at the ground into a tornadic strength vortex \citep[e.g.,][]{parker2023organzied}. For this to occur, several conditions apparently need to be met simultaneously. The sub-tornadic rotation needs be within outflow air that has sufficiently small negative buoyancy so that it does not resist upward acceleration into the low-level updraft \citep{markowski2002direct}. Surface rotation also must experience persistent convergence and stretching. Because this is necessarily below the height of the LFC (where air is either neutrally or negatively buoyant), the bulk of the associated vertical accelerations must be provided by the mesocyclone and its associated dynamic lifting \citep{rotunno1982influence,lilly1986structure,markowski2012pretornadic2,markowski2014influence,coffer2015impacts,coffer2017simulated,goldacker2021updraft}\footnote{In low LCL environments common to tornadic supercells, buoyant vertical pressure perturbation gradients can also have a slight upward contribution to the total acceleration field \citep[see their Fig. 14]{brown2019lcl}.}. This dynamic lifting owes its existence to pressure falls aloft associated with the mesocyclone's circulation. Many studies use the 0 - 1 km vertical perturbation pressure gradient acceleration as a measure of this dynamic lifting provided by the mesocyclone. In addition, tornadogenesis is more likely when there is a vertical alignment of the near-ground, low-level, and mid-level rotation \citep{snook2008effects,guarriello2018effects}, which can be affected by the distribution of shear \citep{markowski2014influence,gray2021midlevel}, the storm-relative flow \citep{brooks1994role,warren2017impact}, properties of the outflow and surges \citep{skinner2014vortex2,marquis2016analysis}, as well as the distribution of negative buoyancy \citep{markowski2017sensitivity} and hydrometeors \citep{loeffler2020differentiating}.  

Within this chain of processes, the point at which a supercell ultimately succeeds or fails in producing a tornado appears is strongly linked to the mesocyclone, specifically the rotation in the lower troposphere \citep{thompson2017damage}. Hence, low-level mesocyclone intensity exerts a substantial influence on tornadogenesis likelihood \citep[e.g.,][]{markowski2014influence,coffer2017simulated,peters2023disentangling}. We will refer to the mesocyclone near cloud-base as the ``low-level'' mesocyclone [in the lower troposphere at approximately 1 km above ground level (AGL)] and consider this level distinct from both the ``mid-level'' mesocyclone further aloft (in the mid-troposphere between $\sim$3-6 km AGL) and the ``near-ground'' rotation that develops much closer to the surface ($<$ 250 m AGL, sometimes referred to as the ``tornado cyclone''). While rigid distinctions between these levels can be somewhat problematic \citep[as discussed in][Section 4c]{markowski2008vortex}, especially at later stages in supercell lifecycles and immediately preceding tornadogenesis, we distinguish the low-level mesocyclone in this way because it is this altitude range where rotation is responsible for the dynamic upward accelerations that must occur below the LFC to produce tornadogenesis.

Current thinking regarding the mesocyclone at low-levels largely originates from the seminal modeling work of \citet{rotunno1985rotation}. By integrating material circuits backward from an area of low-level rotation, \cite{rotunno1985rotation} showed that it was linked to an ``upward tilting of baroclinically-generated horizontal vorticity along the cool air boundary situated upstream of the low-level updraft''. The orientation of the forward-flank baroclinic zone and the storm-relative flow in this region of the storm is such that baroclinically-generated vorticity is predominantly streamwise \citep{klemp1983study}. In simulations with and without rain, rotation in the mid-levels developed consistently through the tilting of environmental horizontal vorticity. On the other hand, low-level rotation was absent in simulations without rain. The \citet{rotunno1985rotation} analysis was performed at the lowest model level (250 m AGL), so there is some ambiguity about whether the ``low-levels'' they described are more applicable to the the low-level mesocyclone rotation near cloud-base (as-in this work) or the origins of surface rotation \citep[as-in][]{davies1993mesocyclogenesis}. 
%Although much of the relevant analysis in \citet{rotunno1985rotation} was performed at the lowest model level (250 m AGL), the development of low-level rotation shown in their material circuits (their Figs. 11, 12) and vortex lines (their Fig. 9) resemble the development of a low-level mesocyclone near cloud base, not the origins of surface rotation (discussed above).  %Instead, \citet{rotunno1985rotation} show circulation developing within the forward flank baroclinic gradient 10-15 minutes prior. %Many modeling studies since have reinforced the importance of storm-generated streamwise horizontal vorticity along the forward flank of supercells to the development of the low-level mesocyclone \citep[e.g.,][]{wicker1995simulation,adlerman1999numerical,beck2013assessment,naylor2014vorticity,marquis2016analysis}. 
In either case, \citet{rotunno1985rotation} has become the defacto reference used to explain the development of low-level mesocyclones \citep[e.g.,][]{markowski1998occurrence,atkins1999influence,wakimoto2000analysis,markowski2003tornadogenesis,shabbott2006surface,orf2017evolution,frank2018entropy,markowski2020intrinsic,fischer2020relative,flournoy2020modes,flournoy2021motion,murdzek2020processes,murdzek2020svc,schueth2021svc,davies2022theory,finley2023svc}. 
%These studies, as well as \citet{rotunno1985rotation}, have varying definitions of `low-level', typically based on the lowest scalar model level (such as 250 m AGL in RK85). In this work, these analyses would generally be referred to as `near-ground', and given the noted importance of baroclinic generation to the development of surface rotation, it is perhaps not surprising to see an outsized role of storm-generated horizontal vorticity in their budgets. 

In one of the most well-observed supercells in history, \cite{markowski2012pretornadic2} used a dual-Doppler wind syntheses to show that as much as 70-90\% of the low-level mesocyclone (at 750 m AGL) was due to storm-generated sources. Other observations also seem to point to forward-flank processes as fundamental to low-level mesocyclones. Circumstantial evidence includes visual cues in low-level cloud features, such as a localized lowering of the cloud base \citep[the ``wall cloud'', often tilted towards the region of forward flank precipitation;][]{fujita1959detailed,atkins2014observations}  due to the influx of precipitation-cooled air into the low-level mesocyclone area. %, inflow bands/beaver tails, and disorganized cloud tags or ``scud'' rising up into the storm (these cloud tags represent locally rain-cooled parcels). 
Additionally, vortex lines around mesocyclones from Doppler radar studies are often configured into `vortex line arches', which are highly suggestive of upward tilting of baroclinically generated horizontal vorticity within the low-level mesocyclone \citep[e.g.,][]{straka2007observational,markowski2008vortex,markowski2012pretornadic1}.
%, as long as the temperature deficit within a storm’s forward flank outflow is not excessively negative \citep{shabbott2006surface}. 
\citet{markowski2010mesoscale} thus summarize the current understanding: ``the formation of low-level mesocyclones usually awaits the development of an extensive forward-flank precipitation region and outflow'' because the ``tilting of the baroclinically enhanced low-level horizontal vorticity produces more significant vertical vorticity at low altitudes than does the tilting of environmental vorticity alone''.

Attention to this concept has been revived by recent interest in the ``streamwise vorticity current'' \citep[SVC; ][]{orf2017evolution}, a localized region of horizontal streamwise vorticity located parallel to the forward flank outflow boundary. Analyzing a very high-resolution simulation of an EF5 tornadic supercell, \citet{orf2017evolution} stated that the SVC was ingested into the storm’s updraft, intensifying the low-level mesocyclone, and leading to the amplification and maintenance of a long-lived violent tornado. This correlation between the intensifying SVC and low-level mesocyclone was expanded upon by \citet{finley2023svc}. While the idea of storm-generated, streamwise horizontal vorticity production within the forward flank has previously been discussed in the literature (as reviewed above), the unprecedented level of detail in these simulations has inspired subsequent exploration in a number of modeling studies \citep{markowski2020intrinsic, schueth2021svc} and observed cases \citep{markowski2018tornadogenesis,murdzek2020svc,murdzek2020processes,schueth2021svc}, and has been an explicit emphasis in recent field projects \citep{weiss2020torus}. 

Even so, it appears that not all supercells have SVCs, and the presence (or lack) of an SVC is not a necessary requirement for tornadogenesis success \citep[or failure;][]{murdzek2020svc}. Even when an SVC is present, it is not guaranteed that the augmented, storm-generated streamwise horizontal vorticity within this feature ends up participating in the low-level mesocyclone \citep{murdzek2020processes}. It seems that substantial uncertainty remains. The present study inherits the question asked by \citet{markowski2012pretornadic2}: ``is large environmental vorticity important, especially at low-levels, because its tilting establishes the base of the mid-level mesocyclone at fairly low elevations?'' 
Indeed, there is accumulating evidence that tornadic environments are distinguished by large \emph{environmental} streamwise vorticity in the lowest 500 - 1000 m AGL \citep{markowski2003characteristics,rasmussen2003refined,miller2006shear,esterheld2008discriminating,nowotarski2013classifying,parker2014composite,coffer2019srh500,coffer2020era5,nixon2022distinguishing}. 
And, a number of modeling studies have attributed the mesocyclone's strong dynamic lifting in the lower troposphere to this environmental source \citep{markowski2014influence,markowski2017sensitivity,coffer2017simulated,goldacker2021updraft,peters2023disentangling}. %Therefore, it seems widely agreed upon that streamwise vorticity from both storm-generated and environmental sources contribute to the low-level mesocyclone rotation, but there is a general ambuigity in how much augmentation is provided by storm-generated sources.  

The vertically integrated storm-relative flux of streamwise vorticity into an updraft, represented by the storm-relative helicity (SRH), is of the greatest dynamical importance in this regard \citep{davies1984streamwise}. SRH must be defined over a layer of some depth, and the choice of this depth is non-trivial. %because the properties of inflow air into supercells strongly influences the potential mesocyclone intensity \citep{peters2023disentangling}. 
Some studies have attempted to define an inflow layer based on thermodynamic properties such that only parcels associated with CAPE and minimal CIN are included \citep[the effective inflow layer or EIL;][]{thompson2007effective,nowotarski2020evaluating}. However, model-based proximity soundings show that SRH very near the ground (e.g., 0 - 500 m AGL; SRH500) is the single most predictive parameter in discriminating significantly tornadic supercells from their nontornadic counterparts, in both United States and European severe weather environments \citep[][]{coffer2019srh500,coffer2020era5}. %; however, for the prediction of tornadoes, the optimal layer has usually been considered over a fixed depth and has progressively been considered over shallower layers, from 0 - 3 km \citep[e.g.,][]{brooks1994environments,rasmussen1998variations} to 0 - 1 km \citep[e.g.,][]{markowski2003characteristics,rasmussen2003refined,thompson2003close} to now layers shallower than 500 m AGL \citep{coffer2019srh500,coffer2020era5,nixon2022distinguishing}. , which could explain the forecast skill of SRH computed in very shallow near-ground layers.
In a pair of simulated supercells (one tornadic, one nontornadic), \citet{coffer2017simulated} showed that the environmental inflow parcels bound for the low-level mesocyclone originated \emph{exclusively below 300 m AGL}. However, the sources of air bound for the low-level mesocyclone across a range of supercells in various environments has not yet been systematically evaluated. 

There is ambiguity in the previous literature concerning the role of environmental versus storm-generated vorticity in the production and maintenance of low-level mesocyclones.  Several prominent conceptual models [e.g., \citet{markowski2008vortex}, their Fig. 19, and \citet{rotunno2017near}, their Fig. 1] seem to depict the low-level and mid-level mesocyclones as sourced from distinct air streams. Some authors have suggested that the low-level mesocyclone is \emph{primarily} attributable to storm-generated baroclinic processes \citep[e.g.,][]{finley2023svc}, while others have argued that the storm-generated vorticity merely \emph{augments} the environmental contribution \citep[e.g.,][]{markowski2010mesoscale}. This uncertainty about the importance of environmental versus storm-generated vorticity in low-level mesocyclone development leads us to our main research questions:

\begin{enumerate}
    \item If low-level mesocyclone-genesis is primarily attributable to storm-generated, forward flank baroclinic generation of horizontal vorticity, then why is near-ground environmental streamwise vorticity such a highly skillful forecast parameter?
    \item How much augmentation, if any, to the low-level mesocyclone from storm-generated horizontal vorticity is necessary to modulate the intensity of low-level dynamic lifting that ultimately can determine whether a supercell fails or succeeds at producing a tornado?
\end{enumerate}

To address these questions, we explore the origins and properties of air parcels that end up in the low-level mesocyclone using a matrix of simulations initialized with wind profiles common to supercell environments observed in nature and representing a spectrum of near-ground horizontal vorticity magnitudes and orientations. 

\section{Methods}\label{methods}

\subsection{CM1 model}

Supercell simulations were performed using Cloud Model 1 \citep[CM1;][]{bryan2002benchmark} release 20.3. Storms were simulated for 3 h on a 200 $\times$ 200 $\times$ 18 km$^{3}$ domain with a horizontally homogeneous environment (described in Section~\ref{methods}b). The inner 100 x 100 km$^{2}$ had a horizontal grid-spacing of 80 m during the period of analysis, while the vertical grid-spacing was stretched from 20 m in the lowest 300 m to 280 m at 12 km. The domain was translated with a constant storm-motion, which was determined iteratively by trial and error to keep the storm approximately centered within the domain. A fifth-order advection scheme, utilizing high-order-weighted essentially nonoscillatory finite differencing, was used with no additional artificial diffusion \citep{wicker2002time}. The subgrid-scale turbulence was parameterized by a 1.5-order turbulence kinetic energy closure scheme similar to \citet{deardorff1980stratocumulus}, with separate horizontal and vertical turbulence coefficients. Open-radiative lateral boundary conditions were employed, and where there was an inward mass flux into the domain, the horizontal winds were nudged back towards the base-state fields along the lateral boundaries (i.e., $nudgeobc = 1$ in the namelist). The upper-boundary had a rigid, free-slip boundary condition, with a Rayleigh damping sponge applied above 14 km, while the bottom boundary condition was semi-slip with a constant surface drag coefficient ($C_d$) of 0.0035 to partially capture frictional effects on within-storm processes. The $C_d$ was calculated using the surface layer scheme by \citet{jimenez2012revised} based on the thermodynamic profile and the mean of the kinematic profiles discussed in Section~\ref{methods}\ref{basestate}. Random perturbations of 0.25 K were added to the initial conditions within the lowest 1000 m AGL. The simulations use a six-category, fully double-moment bulk microphysics scheme from the National Severe Storms Laboratory (NSSL) that explicitly predicts the variable densities of hail and graupel \citep{ziegler1985retrieval,mansell2010sedimentation,mansell2010simulated}. Convection is initialized using the heat flux method of
\citet{carpenter1998entrainment} and \citet{lasher2021entrainment} and is described in more detail below.

\subsubsection{Heat flux convective initialization}

%Many studies have wrestled with the best way to initialize convection in horizontally homogeneous domains, including the classic warm bubble \citep{klemp1978simulation}, low-level convergence of the wind field \citep{loftus2008parameterized}, an updraft nudging tendency in the vertical momentum equation \citep{naylor2012convective}, etc. Test simulations using the updraft nudging technique resulted in intense low-level mesocyclones during the initial development of supercells, regardless of the near-ground wind profile, and was therefore deemed unsuitable for the proposed research questions. Herein, 
We used a Gaussian heat flux based on \citet{lasher2021entrainment} and similar to \citet{morrison2022humidity}, that results in a more natural transition from buoyant plumes to a sustained, mature supercell updraft across all the environments used herein \citep{peters2022shear1,peters2022shear2}. 
%``more realistic transition from buoyant, sheared cloud turrets to a mature supercell''. 
The heat flux was strongest near the surface and exponentially decreased in magnitude radially and vertically. The horizontal width and height of the Gaussian function were 10000 and 2500 m, respectively. The heat flux linearly ramps up from 0 to 2000 W m$^{-2}$ over the first minute of the simulation. This amplitude was maintained for 28 min, then linearly decreased back to zero over 1 min, for a total of 30 min of active heating. In addition to the heating, a very weak Gaussian-shaped forced convergence [$O$(10$^{-4}$)] was applied to the horizontal wind field over the same spatial and temporal dimensions of the surface heating \citep[similar to][]{moser2018cloud}. The combined effect of surface heating and weak convergence resulted in initially shallow thermal-like updrafts that gradually transitioned into a sustained steady supercell updraft \citep{peters2022shear1,peters2022shear2}. 
%Mature updrafts took twice as long to develop with surface heat fluxes compared to the updraft nudging simulations. The footprint of surface heating yielded a gradual buildup of convection, rather than an initial ``bomb'' of latent heating with updraft nudging and the associated feedback on the near-storm environment. 
While the convection organically developed during the first hour of the simulation, the horizontal grid-spacing of the simulation within the inner 100 $\times$ 100 km$^{2}$ was 250 m, which \citet{lasher2021entrainment} found suitable for simulating overturning circulations and entrainment. Afterwards, the horizontal grid-spacing within the inner domain was decreased to 80 m, as discussed next. 

\subsubsection{Downscaling to finer grid-spacing}\label{downscaling}

To adequately resolve features important to the development and maintenance of the low-level mesocyclone and tornadoes, the horizontal grid-spacing within the inner domain of the simulations was downscaled to 80 m an hour into the simulation using the technique of \citet{coffer2022infrasound}. The vertical grid-spacing was unchanged. Downscaling was accomplished by fitting a unique gridded interpolation function (cubic spline) to each two- and three-dimensional array in the model's restart file, essentially creating a one-way nest within CM1.  %While nesting is not common in CM1%\footnote{To the authors knowledge, the only other attempt to modify the CM1 model grid during runtime was by }, there is an extensive body of literature, across multiple numerical model cores, using nests to maximize computational resources and resolution dating back to some of the original three-dimensional simulations of supercells by \citet{klemp1983study}. In fact, in a subsequent release of CM1 (r21.0, April 2022), the model, by default, interpolates restart files to a different model grid for higher (or lower) resolution. 

For comparison, the original 250 m simulations were also run for the full 3 h. Generally the supercells evolved qualitatively similarly between the original 250 m and downscaled 80 m simulations. %\footnote{While not relevant to the current work, the 80 m simulations tended to experience less rightward deviations in the storm motion (Fig.~\ref{hodographs}) and were less isolated than their 250 m counterparts, perhaps due to overly smooth and steady pressure perturbation fields at a lower resolution.}. %, as evidenced by the lack of substantial differences in the time series of maximum vertical velocity and surface vertical vorticity in Fig. 5. 
Subjective comparisons between both resolutions showed that no simulation that produced a tornado-like vortex at 250 m became nontornadic in the 80 m simulation (and vice-versa). The similarities at different resolutions may be somewhat surprising given the low degree of intrinsic predictability within supercells \citep{markowski2017sensitivity,markowski2020intrinsic}; however, it appears, based on the these results, that most of the spread in predictability occurs within the convection initiation and development phase \citep[similar to predictability challenges in operational convective forecasting, e.g.,][]{galarneau2022wofs}. 

\begin{figure*}[t]
\centerline{\includegraphics[width=40pc]{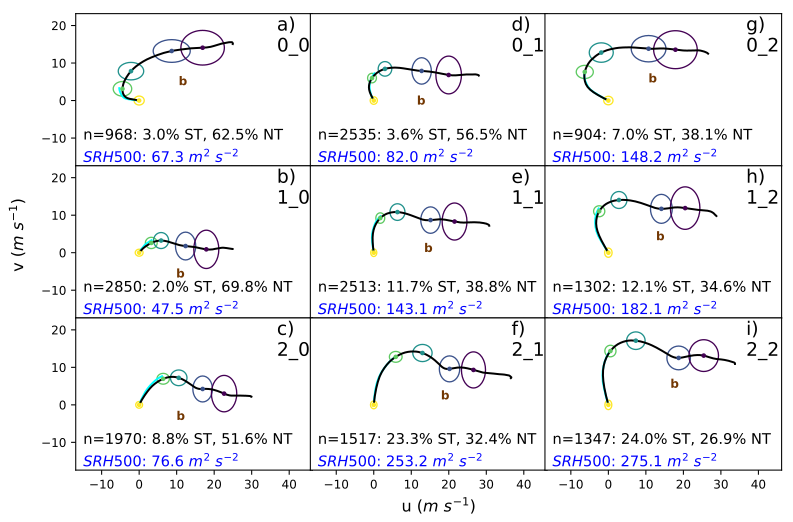}}
\caption{Hodograph diagrams, generated via self-organizing maps, for the nine simulations. The black line is the shifted, steady wind profile used to initialize each simulation from 0 to 12 km, n is the number of samples, ST is the percentage of significantly tornadic cases, and NT is the percentage of nontornadic cases. The cyan line is the shifted original 0 -- 500 m wind profile outputed directly from the SOM. The ellipses are centered at the surface (yellow), 500 m (green), 1 km (turquoise), 3 km (blue), and 6 km (purple), and the radii are equivalent to the standard deviation of the profiles in each node at those points. The Bunkers storm motion, using the original the technique in \citet{bunkers2000predicting} is labeled ``b'' and is used to compute the SRH500 values displayed in each panel. Figure adapted from \citet{goldacker2021updraft}.} \label{hodographs}
\end{figure*}

\subsection{Kinematic and thermodynamic base-state environments}\label{basestate}

To best encompass a range of wind profiles with varying magnitudes of near-ground SRH observed in nature, self-organizing maps (SOMs) were used to identify recurring low-level wind profile patterns from 20,194 model-analyzed supercell proximity soundings previously analyzed by \citet{coffer2019srh500}. SOMs have previously been used to identify recurring features in severe weather datasets \citep{nowotarski2013classifying,anderson2017som,nowotarski2018multivariate,warren2021spectrum,hua2022som}. The SOMs were trained on the 0 - 500 m AGL ground-relative u,v wind components, similar to \citet{goldacker2021updraft}. The 0 -- 500 m AGL layer was selected because it represented the most distinct layer between significantly tornadic and nontornadic supercells in \citet{coffer2019srh500}. All soundings were interpolated to a common vertical grid with 50 m grid-spacing. Training was conducted with three nodes in both the $x,y$ direction, using a learning rate of 0.1 and iterated over 10,000 training steps. %To minimize the total neighborhood sigma values ($\sigma = 0.036$), 
Wind profiles with layers of anti-streamwise vorticity within the lowest 500 m AGL were excluded, leaving a total of 15,906 of cases used in the SOM training. The cases removed represent environments that are not conducive to low-level mesocyclone development and were predominately associated with nontornadic supercells. % \citep[$\sim$12\% of the nontornadic supercell dataset were associated with negative SRH500 compared to less than 1\% in the significantly tornadic supercells in ][]{coffer2019srh500}. 
While SOMs have many uses, the intention herein was to distill a large dataset of environmental soundings into a handful of archetypal hodographs to initialize a reasonable number of supercell simulations for analysis. 

The resulting nine wind profiles from the SOM vary the magnitude and orientation of the near-ground hodograph (Fig.~\ref{hodographs}), generally increasing the magnitude of 0 - 500 m AGL vertical wind shear vector from left to right and shifting the direction of the shear vector from southwesterly to southerly from top to bottom (Fig.~\ref{hodographs}). All profiles have at least 20 m s$^{-1}$ of 0 - 6 km bulk vertical wind shear, supportive of supercells. SRH500 increases from 50 m$^{2}$ s$^{-2}$ in node 1\_0 to 275 m$^{2}$ s$^{-2}$ in node 2\_2. Each ground-relative wind profile from the SOM was subsequently shifted to the origin of the hodograph at the lowest model level (i.e., no surface wind). Shifting the wind profile minimizes the influence of the semi-slip bottom boundary condition on the wind profile over the course of the 3 h simulation without introducing unnatural, or invented, forces into the model's equation set \citep{davies2021invented}. The wind profile was also run through a 1 h CM1 single-column simulation in order to let the profile adjust to the semi-slip bottom boundary condition. Differences were essentially nonexistent between the initial, adjusted, and final far-field wind profile after the 3 h simulation.   %Semi-slip bottom boundary conditions lead to more realistic tornado-like vortices \citep{fischer2022transition}, but their use unavoidably reduces the wind speeds in the near-ground layer in the absence of fully realized turbulent layers and a large-scale pressure-gradient force \citep{markowski2014origins}. Many studies have implemented techniques to minimize this inevitability, including applying the Coriolis force to the perturbation winds \citep[e.g.,][]{roberts2016role,coffer2017simulated,flournoy2020modes}, adding a nudging tendency to the momentum equations \citep{nowotarski2015supercell,markowski2016drag}, and applying a geotroptic wind balance \citep{dawson2019method}. An undesirable side-effect of these methods is the introduction of unnatural, or invented, forces into the model's equation set \citep[the true impact of these fictitious forces is unknown]{davies2021invented}. 

\begin{figure*}[t]
\centerline{\includegraphics[width=25pc]{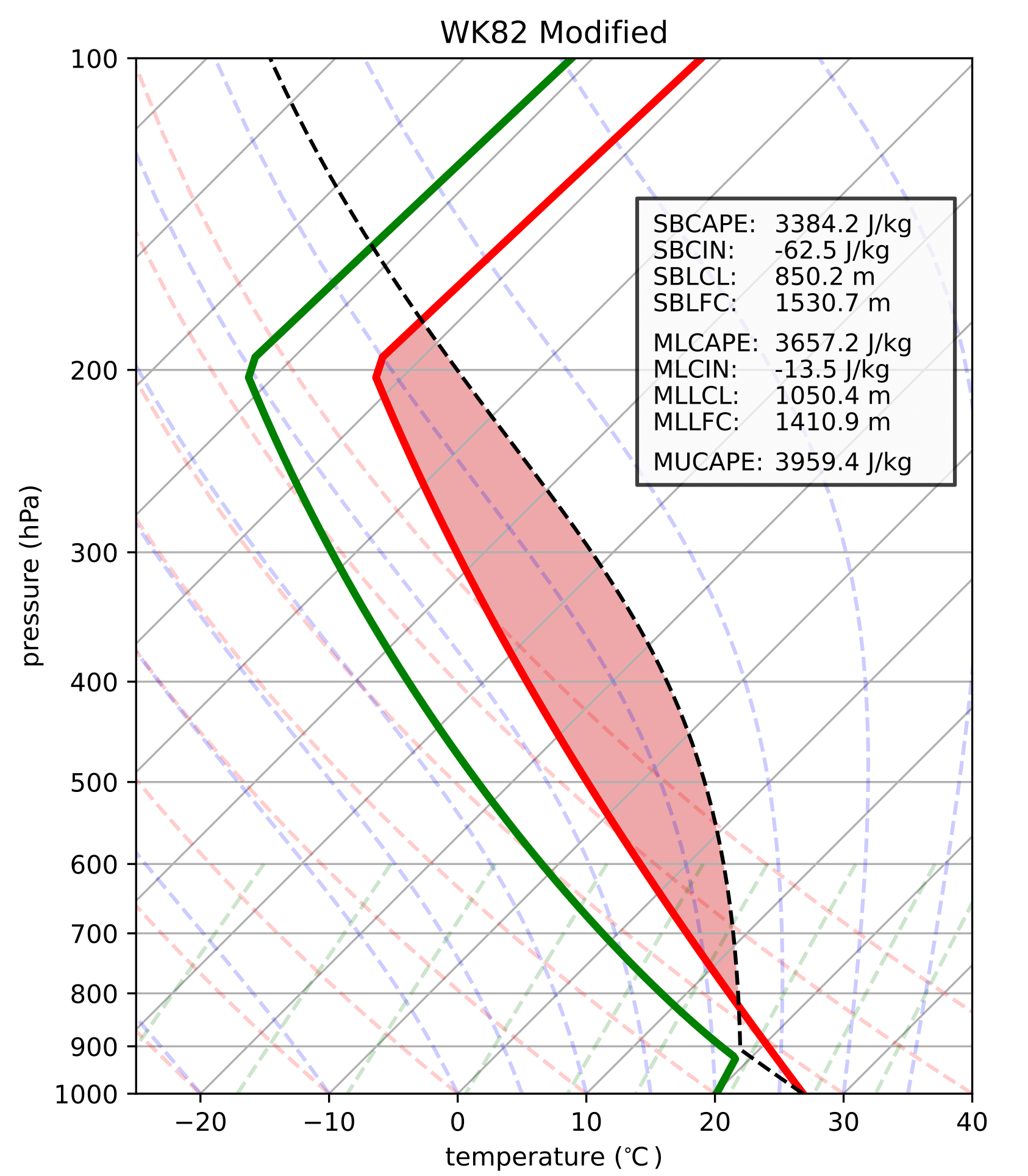}}
\caption{Skew-$T$ log-$p$ diagram of the modified WK82 profile used in each simulation. The environmental temperature $T$ (K) and dewpoint temperature $T_d$ (K) are shown in thick red and thick green, respectively. The solid black line is the pseudoadiabat followed by the mixed-layer parcel, and the area between the temperature curve and the parcel path (CAPE) is shaded in red. Various environmental thermodynamic parameters are also shown.} \label{skewT}
\end{figure*}

Each kinematic profile from the SOM was paired with the same thermodynamic profile (Fig.~\ref{skewT}) in order to isolate the influence of the wind profile on the development of low-level mesocyclones. A modified version of the \citet{weisman1982dependence} sounding was used to generate potential temperature and relative humidity profiles. The modifications are the same as those described by \citet{markowski2020intrinsic} and are designed to minimize the appearance of moist absolutely unstable layers and unwanted/uncontrolled convection initiation.  %This thermodynamic profile is similar to that used by \citet{lasher2021entrainment}, whose surface heat flux initialization we also used. By design the modifications create a overall drier overall throughout the troposphere, which in some sense more accurately reflects typical severe weather environments in the central United States than the original, unmodified profile. 
Mixed-layer CAPE is 3657 J kg$^{-1}$, CIN is -14 J kg$^{-1}$, the LCL is approximately at 1050 m AGL, the LFC is near 1500 m AGL, and the top of the EIL is $\sim$2700 m AGL (Fig.~\ref{skewT}). All parameters were calculated as-in \citet{coffer2019srh500}. These bulk thermodynamic quantities represent the upper-end of the supercell spectrum, and are more similar to the subset of significantly tornadic supercells than to nontornadic supercells in the \citet{coffer2019srh500} dataset.%which has direct effects on the negative buoyancy deficits found within the surface outflow \citep[and indirectly the generation of baroclinic streamwise vorticity production in the forward-flank; ][]{shabbott2006surface}. 

\subsection{Analysis techniques}

\subsubsection{Mesocyclone tracking and definition}

To objectively analyze each supercell, the mesocyclone was tracked over time. The algorithm tracked the peak of a Gaussian, spatially smoothed product of vertical velocity and vertical vorticity (of at least 0.01 s$^{-1}$) at 1 km AGL \citep{werkema2022multistorm}. A temporally smoothed time series of $x,y$ mesocyclone centroids was then computed using a 3rd order polynomial, Savitzky-Golay filter with an 11 point window (i.e., $+/-$ 5 minutes of output). This method reliably tracked the right-moving supercell of interest in each simulation bar one. Simulation 1\_0, which had the smallest near-ground SRH and weakest storm-relative flow, experienced multiple splits and the right-moving storm began to dissipate two hours into the simulation. For this reason, a subjectively defined right moving updraft area was instead defined for the simulation 1\_0 during a key period of interest (i.e., before the storm began to dissipate). 

For analysis purposes, the low-level mesocyclone in each supercell was defined as the grid points at 1 km AGL with vertical vorticity values of at least 0.01 s$^{-1}$ and vertical velocities that exceeded the 90th percentile (Table~\ref{t1}) within a 10 km diameter of the tracked mesocyclone centroids. %and thus the greatest potential for vertical stretching (i.e., largest $\frac{\partial w}{\partial z}$). %This also ensured an equal number of backward trajectories were seeded in each simulation. Finally, the 
In other words, the vertical velocity and vertical vorticity thresholds isolate the portion of the low-level mesocyclone in each supercell that contains up the most intense upward-moving, cyclonically rotating air and thus the greatest potential for vertical stretching \citep[i.e., largest $\frac{\partial w}{\partial z}$, which is highly correlated with dynamic lifting,][]{goldacker2021updraft}.  We chose an altitude of 1 km AGL, near cloud-base for most supercells, not only because mesocyclonic rotation is responsible for upward dynamic accelerations below the LFC (as discussed previously), but also because there is a distinct local maximum in the vertical vorticity field at approximately 1 km AGL across the matrix of simulations presented herein (discussed later). Much of the analysis presented herein was rerun varying the definition of the low-level mesocyclone, including no vertical vorticity requirement, lowering the altitude for what was considered ``low-level'' (i.e., 500 and 750 m AGL), using lower (and higher) thresholds of vertical velocity at 1 km AGL (i.e., the 50th and 99th percentiles), as well as defining the low-level mesocyclone as a coherent area of positive circulation. None of these modifications produced an appreciable change in the overall conclusions.

\subsubsection{Key time periods tornado-genesis/failure}

In order to analyze supercells at similar points in their evolution, a key time period of tornadogenesis or tornadogenesis failure was determined for each simulation. Similar to the definitions of \citet{coffer2017volatility}, vortices at 10 m AGL (i.e., the lowest bottom model level) were considered tornadoes if they met the following criteria: 1) vertical vorticity $\ge$ 0.3 s$^{-1}$ 2) a pressure drop $\le$ -10 hPa throughout the lowest 1 km, and 3) a ground-relative wind speed $\ge$ 29 m s$^{-1}$ (i.e., the EF-0 threshold) within 1 km of the position of maximized Okubo-Weiss (OW) parameter ($OW = \zeta^2 - \left( \frac{\partial u}{\partial y} + \frac{\partial v}{\partial x} \right)^2 - \left( \frac{\partial u}{\partial x} - \frac{\partial v}{\partial y} \right)^2$ ). All three criteria needed to be satisfied for at least two minutes. If these thresholds were not met at any time during a simulation, tornadogenesis failure was defined as the time of maximum OW at 10 m AGL within a 10 km diameter of the tracked low-level mesocyclone centroid point.

\subsubsection{Tracers and backward trajectories}

To visualize the source regions of the low-level mesocyclone, three layers of passive tracers were initialized in CM1 [0 - 500 m AGL, 500 - 1500 m AGL (the approximate height LFC), and 1500 - 2700 m AGL (the top of the EIL)], which were advected within the simulation during integration. % (i.e., near-ground, lower tropospheric but not yet buoyant, lower tropospheric and potentially buoyant). 
The value of the tracer mass mixing ratio in each layer was initially set to 1. % so that the average value of tracer within an area, such as a defined area of updraft, represents the concentration of air from that tracer layer.  
In addition to tracers, backward trajectories were used to determine source regions of the low-level mesocyclone with a finer level of spatial detail than tracers can provide. The method of integrating backward trajectories loosely followed that of \citet{gowan2021downstream}, except that 
%While for most model fields, output was saved every 60 s, the three velocity components were saved every 30 s during the backward trajectory integration time of each simulation. Furthermore, 
between native output intervals (60 s), velocity fields were linearly interpolated into 3 s intervals to ensure that  
% grid-spacing (80 m) divided by the integration time step (3 s) did not exceed $\sim$25 m s$^{-1}$ so that
trajectories do not ``skip'' over entire grid cells during a single integration step. %Although forward trajectories natively within CM1 (integrated at each model time step) are more accurate than backward trajectories \citep{dahl2012uncertainties}, the flow regime of air rising into a low-level updraft should be less susceptible to the large errors associated with highly convergent flows and large velocity gradients near the surface, such as the situation of tracking near-ground tornadic vortex parcels in \citet{dahl2012uncertainties}.
Backward trajectories were initialized at grid points within the defined low-level mesocyclone every 60 s between 5 and 10 minutes prior to the key time period of tornado-genesis/failure ($t_{-10}$ to $t_{-5}$) and were tracked backwards for 30 mins, allowing the final position of the trajectories to be far enough removed from the storm. Given that the timescale for tornado formation is roughly 10 mins \citep{davies2001tornadoes}, the $t_{-10}$ to $t_{-5}$ time window focuses on the point within the low-level mesocyclone's evolution in which the tornadogenesis process is ongoing, not at $t_{0}$ when a tornado has potentially already formed\footnote{Backward trajectories in the $t_{+5}$ to $t_{+10}$ \emph{post-tornadogenesis} time frame have a very similar shape, width, and depth to the pre-tornadogenesis trajectories presented herein. Many of the trajectories in the minutes immediately preceding tornadogenesis ($t_{-2}$ to $t_{0}$) resemble trajectories that result in \emph{near-ground} rotation from \citet{dahl2014imported}, \citet{dahl2015near}, and \citet{fischer2020relative} within downdrafts in the rear-flank outflow before swiftly rising into the
low-level mesocyclone.}. This eludes the potentially problematic issue of determining the exact moment surface rotation should be considered ``tornadic'' \citep[e.g.,][]{houser2022additional}.% The total integration time of 30 mins allowed for the trajectories to be far enough removed from the storm. 

\subsubsection{Material stencils}

In order to address the relative contributions of environmental versus storm-generated horizontal vorticity upon the mesocyclone, we complement the tracer and backward trajectory analysis with forward trajectories within CM1. Using the ``material stencil'' method from \citet{dahl2014imported} and \citet{dahl2023origins}, the initial (or ``imported'') environmental vorticity of a parcel can be separated from the contribution of vorticity generated by the storm. Following \citet{dahl2014imported}, six additional adjacent stencil parcels were initialized surrounding a center parcel at distances of 0.5 m. Over time, the stencils are deformed, and the embedded initial vorticity vector is reoriented accordingly (behaving as a material fluid that is tilted and stretched). Thus, for parcels that subsequently enter the low-level mesocyclone, the component of the vertical vorticity that is due to the initial ambient vorticity can be derived based on the final configuration of the stencil. The storm-generated component is simply the residual between the known vorticity at some final time and the rearranged initial vorticity component.

Since the initial parcel locations may not be representative of completely undisturbed base-state air due to far-reaching storm influences on the environment \citep[e.g.,][]{parker2014composite,wade2018comparison,coniglio2020insights}, we extended the stencil method from \citet{dahl2014imported} to parse out two components of initial vorticity. The initial stencil vorticity vectors include both the base-state values from model initialization ($\vec{\omega}_{env}$) and any perturbations that have developed between the model start time and the stencil initialization time ($\vec{\omega}_{pert}$). Therefore the final vertical vorticity ($\zeta$) of the low-level mesocyclone ($LLM$) is

\begin{equation}
    \zeta_{LLM} = \zeta_{I_{env}} +  \zeta_{I_{pert}} + \zeta_{SG}
\end{equation}

where $I$ is the component of the initial stencil vorticity that is rearranged by the storm via tilting and stretching, divided into the base-state environment ($env$) and pre-existing perturbations ($pert$), and $SG$ is the residual storm-generated component that contains all the nonconservative vorticity production processes, such as baroclinic, subgrid-scale mixing, and diffusion (and the eventual rearrangement of those nonconservative processes). Both base-state environment ($env$) and pre-existing perturbations ($pert$) are treated as `frozen to the flow' and can be rearranged via tilting and stretching. Because it is unknown whether or not $\vec{\omega}_{pert}$ represents a prior process that represents a rearrangement of the initial environmental vorticity vector or vorticity produced by the storm itself,
%stretching of $BT$ horizontal vorticity of environmental inflow air accelerating towards the storm or prior storm-generated $NBT$ vorticity in the far-field, 
the three components of $\zeta_{LLM}$ will be reported separately. The total of $\zeta_{I_{env}}$ and $\zeta_{I_{pert}}$ always sums to $\zeta_{I}$, which is what was originally presented in \citet{dahl2014imported}. We hope that separating the base-state and perturbation vorticity vectors in this manner helps to address historical concerns that analyzed trajectories were potentially not fully removed from the storm's influence at the time of initialization. 

Forward trajectories were seeded within model restart files 40 minutes before the key time period of tornado-genesis/failure for each simulation and integrated forward natively within CM1 for 35 minutes. This results in parcels with at least 30 minutes of output history during the same five minute composite period ($t_{-10}$ to $t_{-5}$) highlighted by the backward trajectories. Horizontally, parcels were launched upstream of the low-level mesocyclone within a unique horizontal bounding box for each simulation encompassing an estimated 75\% of the low-level mesocyclone inflow area based on the backward trajectories, with a 2 km buffer upstream in the $+x$, $+y$, and $-y$ directions (to account for potential errors in backward trajectories). Vertically, parcels were defined over a 1500 m layer starting at 30 m AGL (i.e., the second lowest model level\footnote{The second lowest model level was the lowest chosen because surface drag always opposes the local wind field at the lowest model grid point with a  ``semi-slip'' bottom-boundary condition, yielding questionable horizontal vorticity fields \citep[i.e.,][]{wang2020memory,wang2023modeling}.}), except for the 1\_0 simulation where parcels were extended up to 2000 m due to higher parcel origin heights (shown later). The center stencil parcels were defined on an isotropic 100 m grid within the bounding box, with the six adjacent stencil parcels initialized surrounding each center parcel at distances of 0.5 m. Each simulation had between 2,000,000 and 12,000,000 total forward trajectories, depending on the areal extent of the inflow and thus the size of the bounding box (shown later). Parcel data were saved every 15 s and low-level mesocyclone trajectories were identified using the same thresholds as the backward trajectories, i.e., vertical velocity greater or equal to the 90th percentile (Table~\ref{t1}) and at least 0.01 s$^{-1}$ of vertical vorticity (within $+/-$ 10 m of 1 km AGL).%Overall, our goal with the forward trajectories is to identify the dominant contribution of vertical vorticity to the low-level mesocyclone over the majority of the inflow air. The seeded locations do not represent an exhaustive source of the inflow due to computation constraints. In particular, a 100 m vertical parcel grid biases the results by only seeding a handful of vertical levels within the 0 - 500 m AGL layer where a majority of the parcels that contribute to low-level mesocyclone originate from (and often posses the highest horizontal streamwise vorticity values). On the other hand, the bounding box ecludes parcel locations outside of the 75\% KDE, which often represents forward flank parcels (Fig. 11).  

\section{Results}

\subsection{General characteristics of the simulations}

\begin{figure*}[t]
\centerline{\includegraphics[width=40pc]{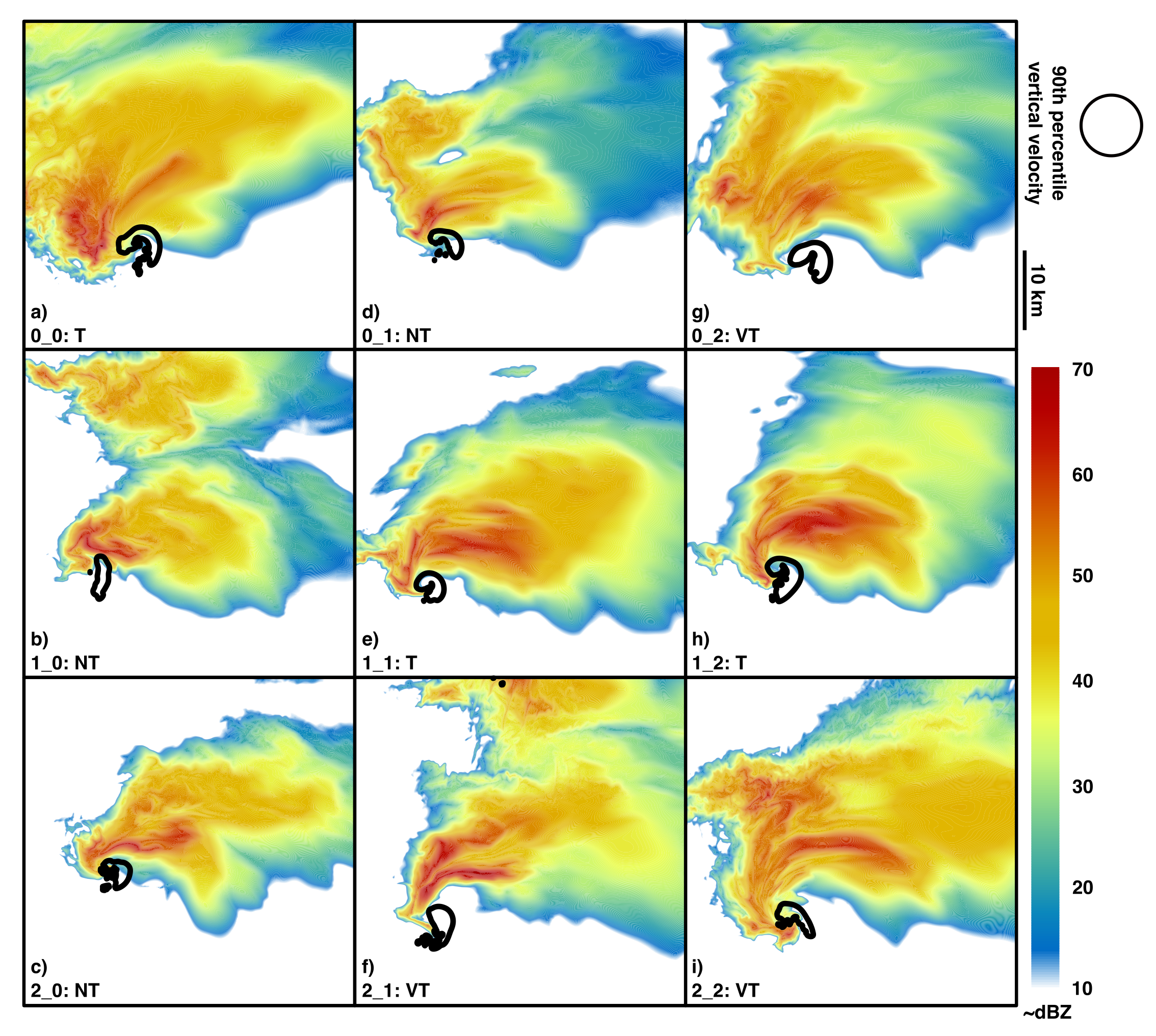}}
\caption{Horizontal cross sections of 10 m AGL reflectivity (dBZ; shaded) at the key time period of tornado-genesis/failure (see Table~\ref{t1}) for each simulation and the 90th percentile of vertical velocity associated with the updraft at 1 km (black contour). Each panel is labeled nontornadic (NT), tornadic (T), or violently tornadic (VT).} \label{reflectivity}
\end{figure*}

\begin{figure*}[t]
\centerline{\includegraphics[width=40pc]{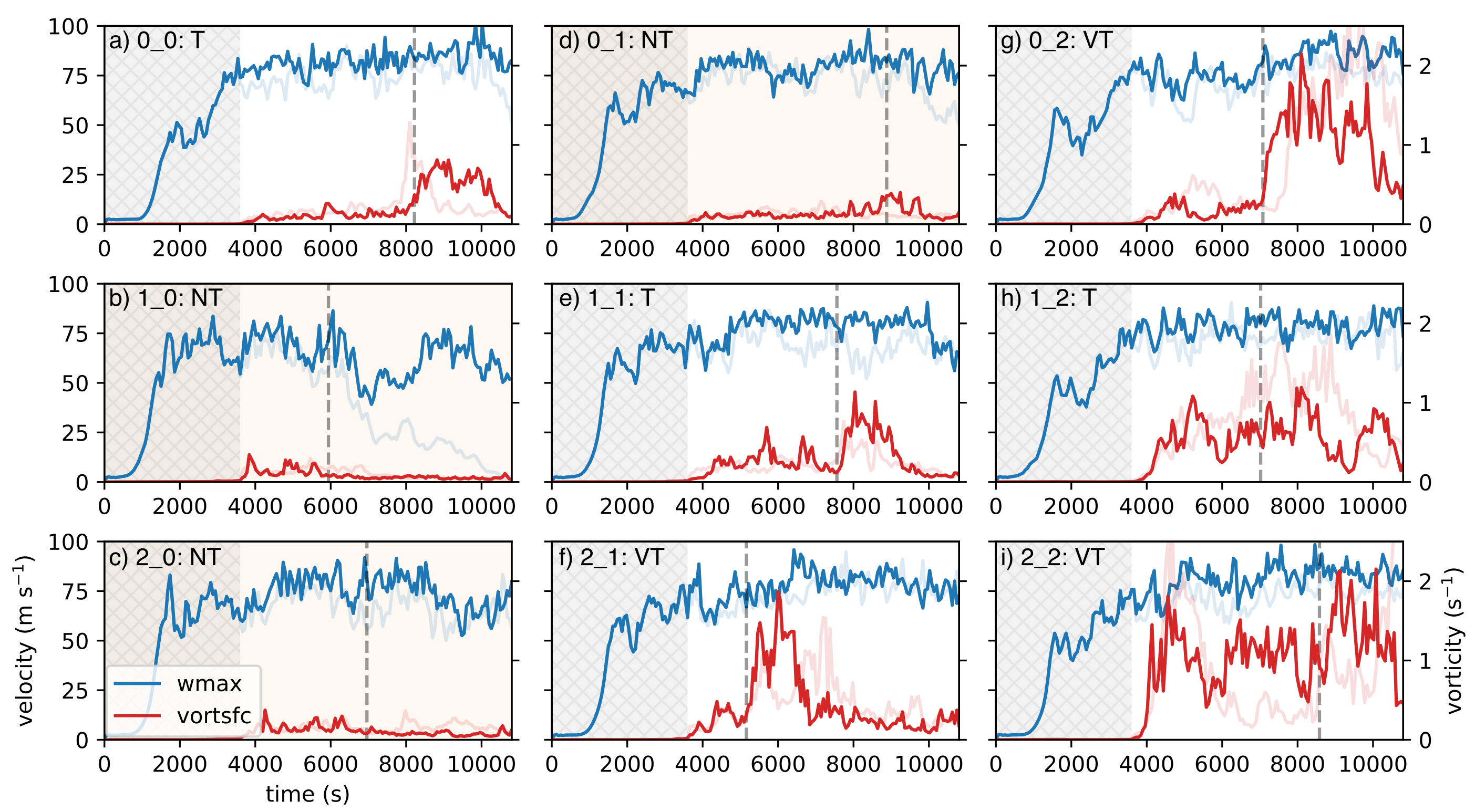}}
\caption{Time series' of maximum domain $w$ (blue, m s\textsuperscript{-1}, left $y$ axis) and maximum surface $\zeta$ (red, s\textsuperscript{-1}, right $y$ axis), with the pre-downscaling period hatched. The key time period for each simulation is shown as a vertical dashed line (note the key time period for nontornadic supercells is not necessarily at the time of peak vertical vorticity in the domain). Also shown as faded lines are the same time series from original 250 m simulations ($\zeta_{sfc_{250m}}$ is scaled to account for differences in magnitude due to lower resolution). Each panel is labeled nontornadic (NT), tornadic (T), or violently tornadic (VT). Panels faded yellow correspond to NT simulations. } \label{timeseries}
\end{figure*}

All nine wind profiles from the SOM %, combined with a favorable thermodynamic profile, 
resulted in supercellular convection for the majority of the 3 hour simulation time. Eight of the nine simulations develop quasi-steady, right-moving supercells with persistent (and trackable) low-level mesocyclones of varying intensity (Fig.~\ref{reflectivity}). The lone exception is simulation 1\_0, which due to the lack of curvature and storm-relative flow in the hodograph (Fig.~\ref{hodographs}b), experiences a succession of splits. After two hours, only disorganized multi-cell convection exists throughout the domain in simulation 1\_0 and peak vertical velocities drop off substantially compared to the other eight simulations (Fig. ~\ref{timeseries}a). Despite the unsteady nature, at times, the southern-most, right-moving storm in simulation 1\_0 periodically displays supercellular features in the reflectivity field, such a hook echo\footnote{The most prominent of such instances is chosen as the key time period and the centroid of the low-level updraft (albeit weaker than any other simulation; Table~\ref{t1}) was manually defined over time.} (Fig.~\ref{reflectivity}b). Due to the lack of persistent low-level mesocyclone, the 1\_0 storm is probably of less interest to the supercell tornadogenesis problem; however, for completeness, we have not excluded any of the analyses for this simulation. In contrast to 1\_0, the other eight simulations experience relatively steady maximum vertical velocities of over 70 m s$^{-1}$ from 1 hour onwards (when the downscaling occurred; Fig.~\ref{timeseries}), as the initial convection produced by the surface heat flux initialization coalesces into singular, dominant updraft.  

By happenstance, the nine simulations can be evenly separated into three groups of three, nontornadic (1\_0, 2\_0, 0\_1), tornadic (0\_0, 1\_1, 1\_2), and violently tornadic (2\_1, 0\_2, 2\_2). The tornadic simulations are found further to right of the SOM (Fig.~\ref{hodographs}) following trends in the increasing magnitude of SRH500. The three nontornadic simulations all had SRH500 less than 100 m$^2$ s$^{-2}$, although this threshold did not preclude the 0\_0 simulation from becoming tornadic (indicating that maybe some amount of storm-generated augmentation was more prominently present in this simulation). Qualitatively, the nontornadic supercells display muted trends in surface vertical vorticity (Fig.~\ref{timeseries}b-d) and never meet the threshold of a deep, long-lasting vortex underneath the main low-level mesocyclone, whereas the tornadic simulations experience abrupt jumps in the maximum surface vertical vorticity (Fig.~\ref{timeseries}a,e-i). The six tornadic simulations can be further delineated by their max $OW$. Vortex intensity increases from $OW \sim$ 0.2 in three of the simulations (0\_0, 1\_1, 1\_2) to $OW > 1$ in the violently tornadic ones (2\_1, 0\_2, 2\_2; Table~\ref{t1}). Some of the tornadic simulations experience multiple periods of tornadic activity. In such cases, our analysis on the origins of inflow into the supercells is performed on the low-level mesocyclone that resulted in the most intense tornado, as defined by $OW$, throughout the simulation.  

\begin{figure*}[t]
\centerline{\includegraphics[width=40pc]{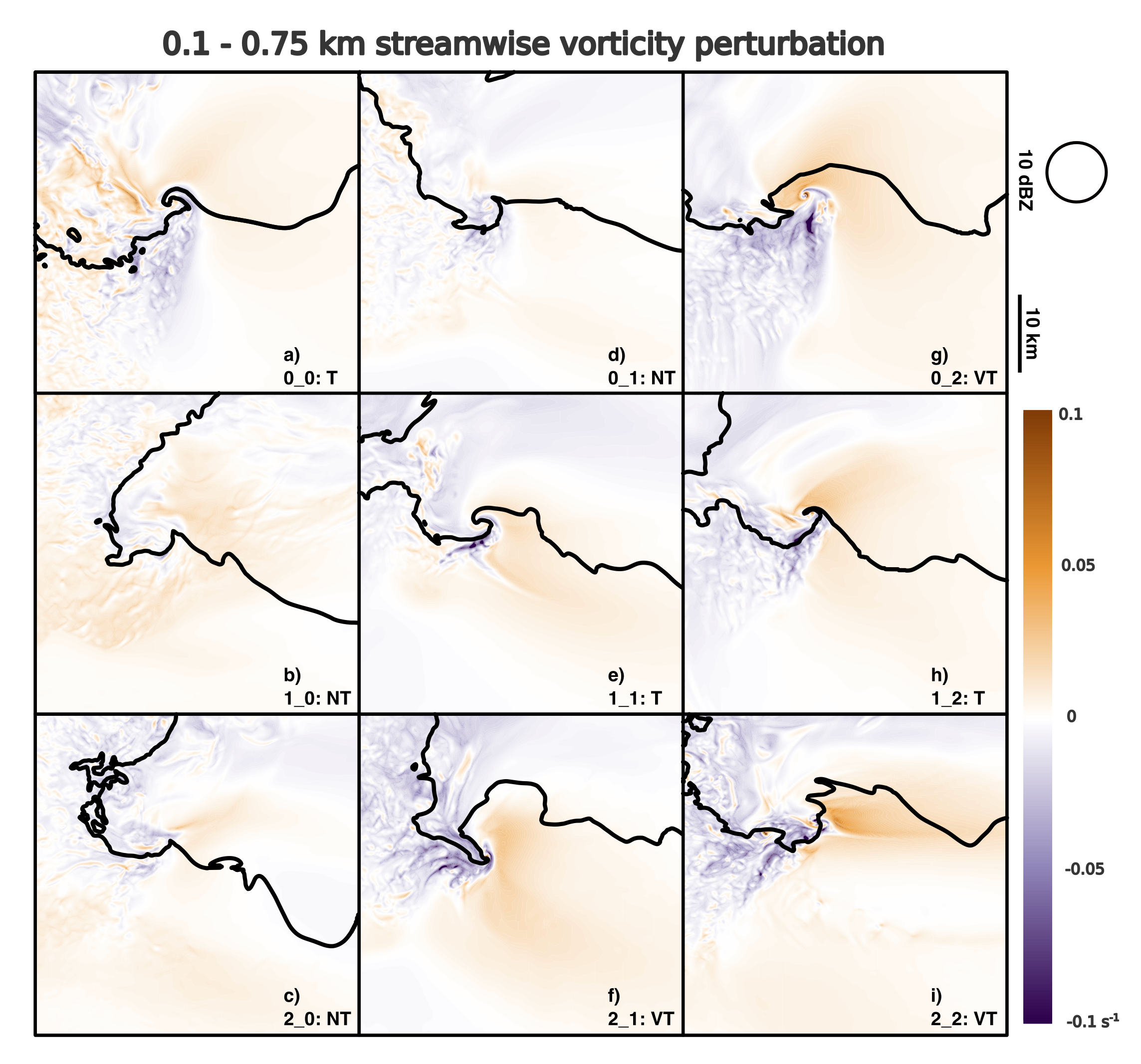}}
\caption{Horizontal cross sections of the layer averaged streamwise vorticity perturbation between 0.1 and 0.75 km AGL (s$^{-1}$; shaded) at the key time period of tornado-genesis/failure (see Table~\ref{t1}) for each simulation and the 10 dBZ reflectivity outline (black contour). Each panel is labeled nontornadic (NT), tornadic (T), or violently tornadic (VT).} \label{SVC}
\end{figure*}

Each supercell, regardless of tornadic outcome, displays a region of enhanced streamwise vorticity relative to the base-state environment throughout the near-inflow and within the forward flank region at the key time period of tornado-genesis/failure (Fig.~\ref{SVC}). Whether or not these features would be classified as SVCs is beyond the scope of this paper, but there is a correlation between the strength of the low-level updraft (Table~\ref{t1}) and larger streamwise vorticity perturbations in the inflow and forward flank regions (Fig.~\ref{SVC}f-i). The cause of this correlation is unclear. It is possible that enhanced streamwise vorticity leads to stronger updrafts. It is also possible that stronger low-level updrafts induce greater horizontal stretching of streamwise vorticity via near-ground horizontal accelerations. The latter explanation leads to the most intense regions of streamwise vorticity within the SVC vorticity budgets analyzed by \citet{schueth2021svc}. Nevertheless, this question regarding the importance the enhanced regions of streamwise vorticity further motivates the subsequent analysis of the origins of inflow and vorticity within supercell low-level mesocyclones, which we explore next. 

\begin{table*}[t]
\caption{Summary of characteristics from each simulation, including the defined key time period of tornado-genesis/failure (min), the threshold of the 90th percentile of vertical velocity (m s$^{-1}$) at 1 km AGL within a 10 km diameter of the low-level mesocyclone centroid at the key time period, the maximum 1 km updraft speed (m s$^{-1}$) at the key time period, the maximum ground-relative wind speed (m s$^{-1}$) at 10 m AGL within 1 km of the position of maximized Okubo-Weiss (OW) parameter for tornadic/nontornadic period of interest (and the corresponding EF rating), the maximum vertical vorticity (s$^{-1}$) for tornadic/nontornadic period of interest, the maximum OW for tornadic/nontornadic period of interest (s$^{-1}$).}\label{t1}
\begin{center}
\begin{tabular}{c|cccccc}
\hline\hline
 & \begin{tabular}{c}key time \\ period\end{tabular} & \begin{tabular}{cc}90th percentile\\ 1 km updraft\\ threshold\end{tabular} & \begin{tabular}{cc}max 1 km\\ updraft speed\end{tabular} &\begin{tabular}{c}max tornadic \\  surface wind speed\end{tabular} & \begin{tabular}{c}max tornadic \\ surface $\zeta$\end{tabular} & \begin{tabular}{c}max tornadic \\ surface $OW$\end{tabular}\\
\hline
 0\_0  & t = 137 & 8.7 & 20.5 & 51.6 (EF2) & 0.81 & 0.21\\
 1\_0  & t = 99 & 3.3 & 7.6 & nontornadic & -- & --\\
 2\_0  & t = 116 & 6.0 & 15.1 & nontornadic & -- & --\\
 0\_1  & t = 148 & 6.9 & 13.1 & nontornadic & -- & --\\
 1\_1  & t = 126 & 6.9 & 21.5 & 52.8 (EF2) & 1.13 & 0.29\\
 2\_1  & t = 86 & 10.1 & 27.9 & 84.7 (EF4) & 1.87 & 1.01\\
 0\_2  & t = 118 & 8.4 & 29.3 & 96.6 (EF5) & 2.15 & 1.20\\
 1\_2  & t = 117 & 6.6 & 30.5 & 67.9 (EF3) & 1.19 & 0.22\\
 2\_2  & t = 143 & 14.6 & 41.6 & 97.2 (EF5) & 2.15 & 1.32\\

\hline
\end{tabular}
\end{center}
\end{table*}

\subsection{Origins of low-level mesocyclone inflow air}

Supercells are largely considered to be products of the environments in which they form. The vertical distribution of quantities such as temperature, moisture, and winds, exert substantial influence over a supercell's evolution. While previous studies have advanced our understanding of the inflow properties that favor supercells that produce tornadoes (compared to seemingly similar nontornadic supercells), questions still remain about where, both horizontally and vertically, supercells of varying intensity source most of their inflow air into the low-level updraft and mesocyclone. This in turn may shed light on the comparative importance of environmental air versus air that is modified within the storm.

\subsubsection{Tracers}

\begin{figure*}[t]
\centerline{\includegraphics[width=20pc]{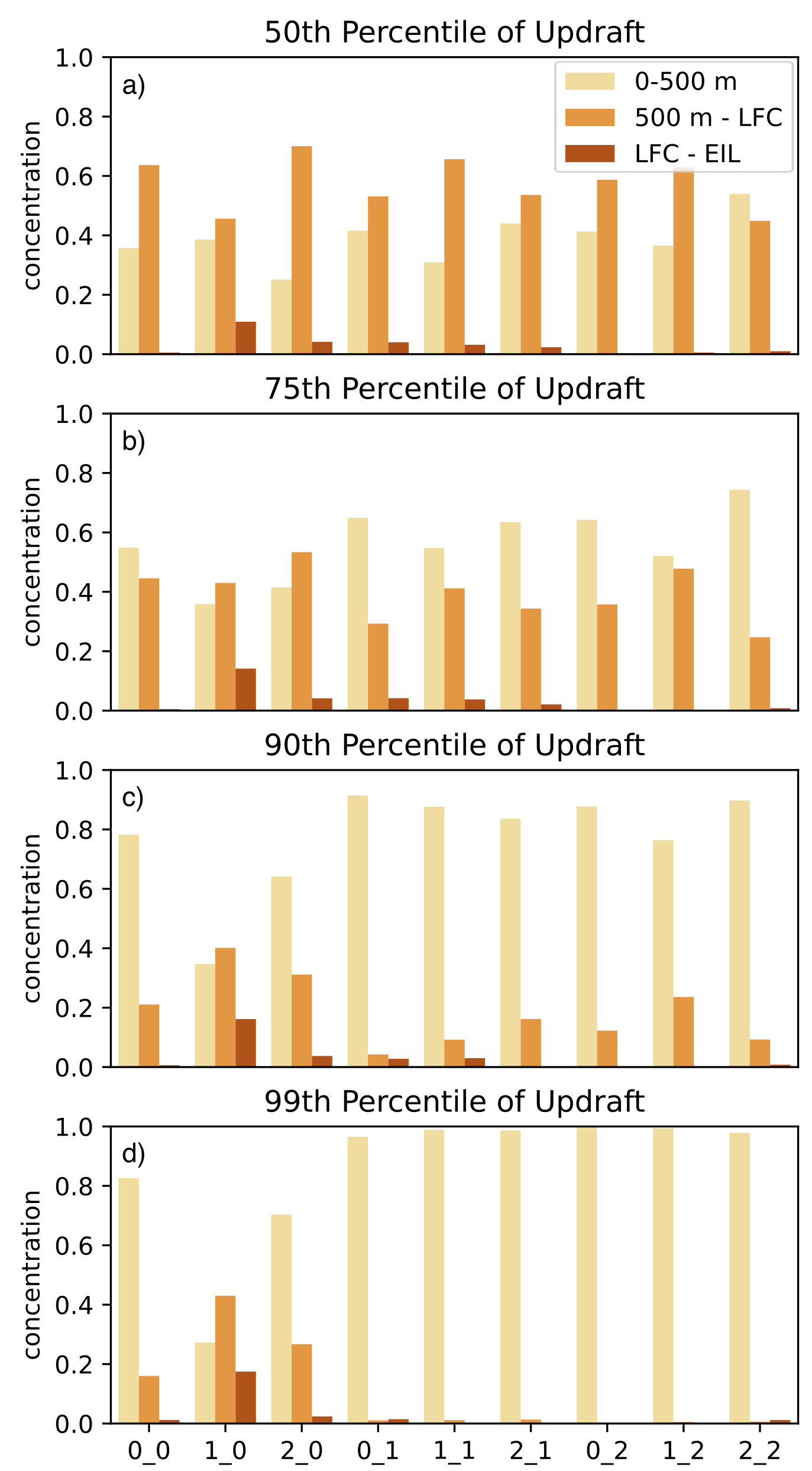}}
\caption{Average tracer concentration in the 0--500 m AGL, 500 m -- LFC, and the LFC--EIL layers within a 10 km diameter of the 1 km AGL low-level mesocyclone centroid within an updraft area at and above various vertical velocity thresholds (50, 75, 90, and 99th percentile) during a five minute composite period ($t_{-10}$ to $t_{-5}$) relative to the key time period of tornado-genesis/failure.} \label{tracer_percent}
\end{figure*}

To first broadly understand the height from which the low-level updraft draws most of its inflow, we use tracers initialized within three layers: 0 -- 500 m AGL, 500 m -- LFC, and the LFC -- EIL. Within a 10 km diameter of the 1 km AGL low-level mesocyclone centroid, the progressively stronger parts of the low-level updraft are increasingly made up of air originating from below 500 m during a five minute composite period ($t_{-10}$ to $t_{-5}$) relative to the key time period of tornado-genesis/failure (Fig.~\ref{tracer_percent}). For the 50th percentile of vertical velocity ($\leq$ 2 m s$^{-1}$) and above, the supercell's low-level updraft is a mixture of air from below the LFC, with the largest proportion of air generally originating between 500 m AGL and the LFC (Fig. ~\ref{tracer_percent}a). As the updraft threshold is increased (50th$\rightarrow$ 75th$\rightarrow$ 90th$\rightarrow$ 99th percentile) from gently rising air to the fastest rising air (and thus the largest $\frac{\partial w}{\partial z}$), the concentration of air from the near-ground layer increases substantially across seven out of the nine simulations. At and above the 99th percentile of updraft values within the low-level mesocyclone, most simulations contain essentially pure, undiluted near-surface air, with a concentration of tracer mass mixing ratio of nearly 1 (Fig.~\ref{tracer_percent}d), especially in the tornadic simulations. While tracers cannot indicate whether this near-ground air is coming directly from the environment or has passed through the forward flank baroclinic zone, concentration values approaching unity within the core of the low-level updraft are indicative of air that is largely unmodified by the storm's outflow since re-ingested forward flank air parcels would tend to experience dilution from mixing. Simulations 1\_0 and 2\_0 (both nontornadic and having the lowest environmental SRH500 values) are noticeable outliers to this trend, although even in 2\_0, at least 60\% of the low-level updraft air is being fed by the near-ground layer. Compared to other tornadic simulations, the 0\_0 supercell has marginally less ($\sim$15--20\%) near-ground air within the low-level mesocyclone (an indication that this low-level mesocyclone, in a lower SRH500 environment, perhaps is supplemented by other sources of air). 

The explanation for the importance of the near-ground layer in the low-level mesocyclone relates to the ascent angle ($\phi$) of the storm-relative wind into an updraft. As inflow air approaches the storm and enters the footprint of the updraft, vertical tilting of horizontal vorticity occurs. From \citet{peters2023disentangling}, the tilting of horizontal vorticity in a perfectly streamwise environment can be expressed as:

\begin{equation}
\label{eq:peterseq}
\zeta_{LLM} = \omega_s \frac{w}{V_{SR}} = \omega_s  \tan \phi ,
\end{equation}
where $\omega_s$ is streamwise vorticity, $w$ is the vertical velocity of the updraft, and $V_{SR}$ is the storm-relative wind [\citet{davies2022theory} presents a similar equation]. Although the wind profiles in Figure~\ref{hodographs} are not perfectly streamwise, Eq.~\ref{eq:peterseq} represents a good first order approximation of the tilting of horizontal vorticity into a mature, right-moving supercell's updraft.

The simplest interpretation of Eq.~\ref{eq:peterseq} is that the tilting of $\omega_s$ is related to the slope of trajectories entering an updraft (i.e., ``rise over run'' or $\frac{w}{V_{SR}}$).  The efficiency of the tilting of environmental streamwise vorticity into the vertical is modulated by the balance between vertical and horizontal motion \citep[also discussed in][]{davies1984streamwise,droegemeier1993influence}. \citet{peters2023disentangling} found that $\phi$ varies with updraft width and storm-relative flow, but the median $\phi$ for parcels bound for the low-level mesocyclone across a spectrum of supercell wind profiles was approximately 10$^{\circ}$, averaged across all their simulations. Thus, the slope of the trajectories into low-level mesocyclones is fairly gentle. Only parcels originating from the near-ground layer are likely to have fully converted $\omega_s$ into $\zeta$ by 1 km AGL. Thus, as suggested by \citet{markowski2012pretornadic2}, large near-ground streamwise vorticity establishes the base of the low-level mesocyclone as close to the surface as possible given typical ascent angles, inducing a ``dynamical feedback'' process of pressure falls and upward directed perturbation pressure gradient accelerations \citep{goldacker2021updraft} needed for lifting and stretching negatively buoyant, circulation-rich air within the supercell's outflow. 

\begin{figure*}[t]
\centerline{\includegraphics[width=35pc]{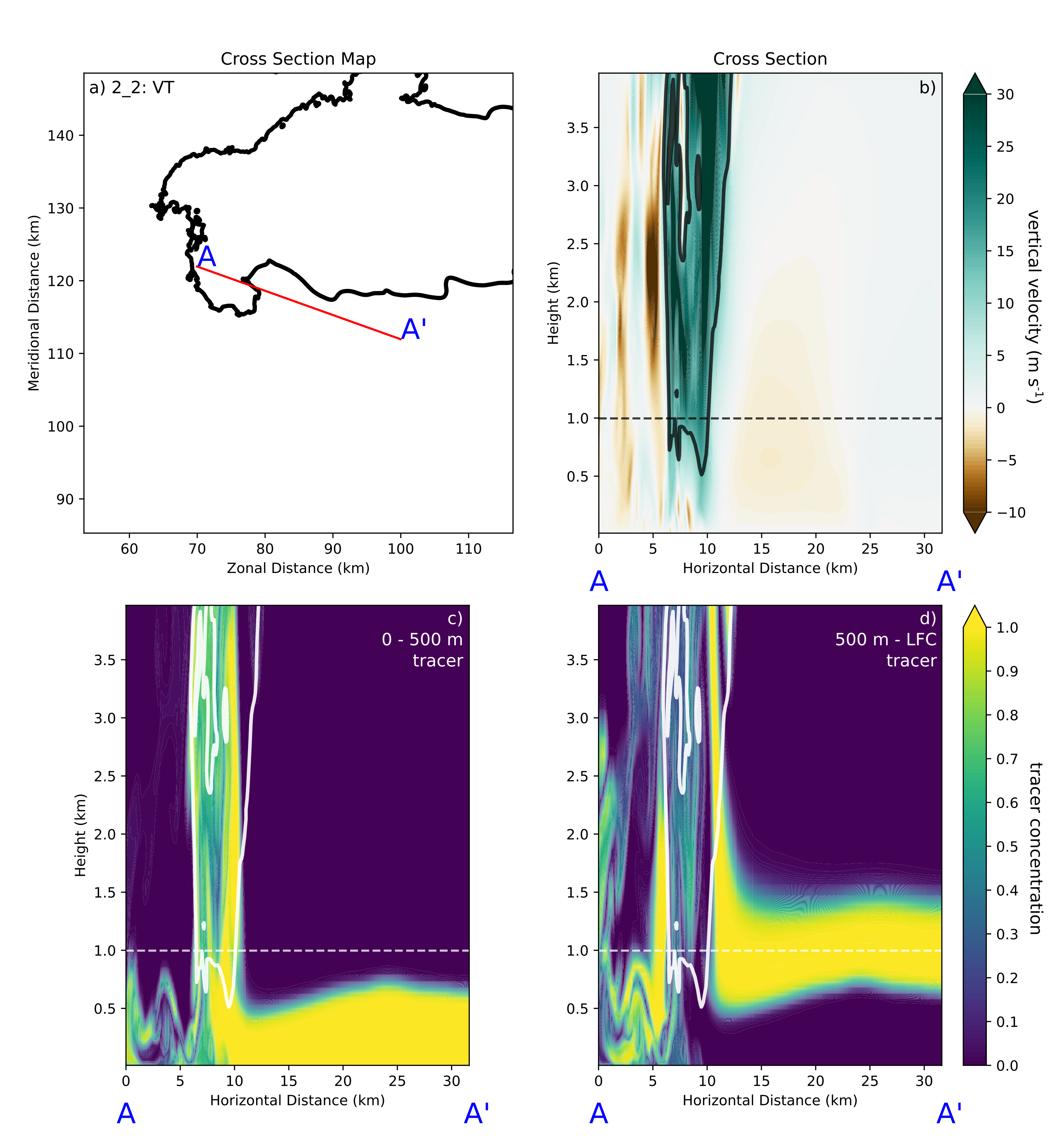}}
\caption{a) Horizontal cross section of 10 dBZ reflectivity for the 2\_2 violently tornadic (VT) supercell at the key time period of tornadogenesis. Path of vertical cross-section used in subsequent panels is shown in red from A to A$^{'}$. b) Vertical cross-section of vertical velocity (shaded) with 1 km AGL labeled as a dashed line and 15 m s$^{-1}$ vertical velocity contour in black. c) As-in b), except shaded is the 0 -- 500 m tracer and the 15 m s$^{-1}$ vertical velocity contour is in white. d) As-in c), except shaded is the 500 m -- LFC tracer.\\} \label{crosssection}
\end{figure*}

In the current simulations, cross-sections through the low-level updraft display similar trends in ascent angles and tracer concentrations within the supercell's updraft at varying altitudes. For the 2\_2 supercell (Fig.~\ref{crosssection}), even prior to tornadogenesis, an intense core of vertical velocities greater than 15 m s$^{-1}$ extends down to 500 m AGL (Fig.~\ref{crosssection}b), with the maximum 1 km updraft exceeding 40 m s$^{-1}$ (Table~\ref{t1}). Within the updraft from 500 m -- 1 km AGL, the concentration of air from the near-ground layer is essentially one (Fig.~\ref{crosssection}c). In fact, this is generally the case within the core of the updraft up to 2 km AGL. At this point, a much larger concentration of air from 500 m to the LFC is present (Fig.~\ref{crosssection}d). Both \citet[their Fig. 5]{nowotarski2020evaluating} and \citet[their Fig. 14]{lasher2021entrainment} show  examples of this gentle ascent layer, where air in the upper part of the inflow layer (i.e., 500 m -- 2 km AGL) does not contribute to the core of the updraft until much farther aloft (i.e., 2 -- 4 km AGL). Below 2 km AGL, what little air that is present from above 500 m is predominately found along the downshear (i.e., the eastern) edge of updraft (Fig.~\ref{crosssection}d). 

\begin{figure*}[t]
\centerline{\includegraphics[width=40pc]{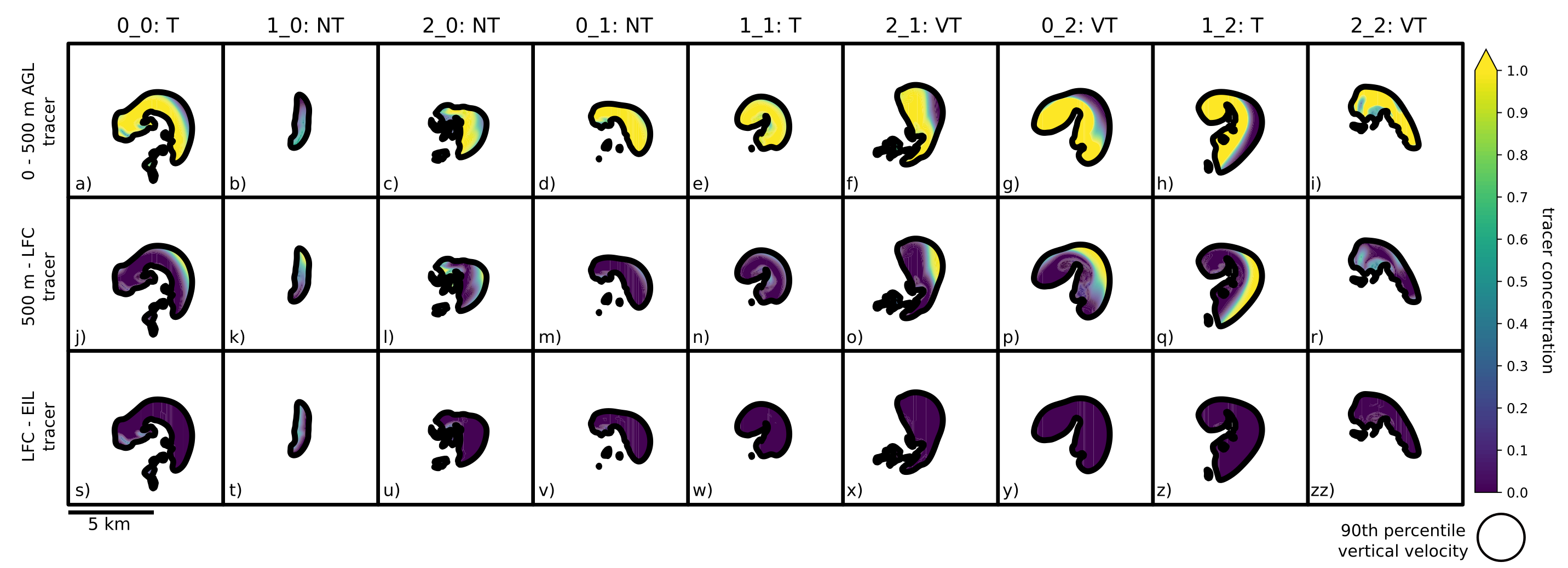}}
\caption{Horizontal cross sections of tracer values for each simulation within the 90th percentile of updraft at 1 km (black contour) at the key time period of tornado-genesis/failure (see Table~\ref{t1}) for (top) 0--500 m AGL, (middle) 500 m -- LFC, and (bottom) the LFC--EIL tracer layers. Each column is labeled nontornadic (NT), tornadic (T), or violently tornadic (VT).} \label{tracer_spatial}
\end{figure*}

Across eight of the nine simulations herein (with the transient 1\_0 supercell again being the outlier), for updrafts defined by the 90th percentile of vertical velocity and above, the highest concentration of air is definitively from the 0 -- 500 m layer (Fig.~\ref{tracer_spatial}a,c-i). Where air above this layer contributes most substantially is along the eastern, downshear flank of the updraft (Fig.~\ref{tracer_spatial}j,l-q). This is consistent with this air stream not yet being fully titled into the vertical (Eq.~\ref{eq:peterseq}) and residing along the edge of the updraft footprint. Air originating from above the LFC is not found with any consistency within the core of the supercells' updraft at any height within the troposphere (not shown), and accordingly is virtually non-existent in the low-level mesocyclone (Fig.~\ref{tracer_spatial}s-zz). Although this is likely not surprising given the altitude of the LFC ($\sim$ 1.7 km), there is a historical precedence in tornado forecasting of integrating SRH over depths much greater than the LFC [e.g., 0 -- 3 km AGL SRH in \citet{rasmussen1998baseline} and SRH in the EIL (ESRH) in \citet{thompson2007effective}]. %, which implicitly implicates a participation of air parcels in the low-level mesocyclone within this layer.
%a correlation between the participation of air parcels within this layer to a storm's ability to produce tornadoes. 
While shallower layers of SRH have the highest correlation with low-level updraft and mesocyclone intensity \citep[compared to the mid-level mesocyclone;][]{peters2023disentangling}, the height at which the wind profile no longer affects tornado potential is currently unknown. Any statistically significant differences in wind profiles between nontornadic and tornadic supercells above the lower troposphere could be due to direct influences of the mid-level updraft/mesocyclone at lower altitudes \citep[such as lowering the base of the mid-level mesocyclone, as suggested by][]{markowski2012pretornadic2} or indirect influences on the storm \citep[such as modifying the deviant rightward storm motion or altering the downstream distribution of hydrometeors relative to the updraft, as suggested by][]{coniglio2020insights,coniglio2022mesoanalysis}. However, in the simulations presented herein, air above the LFC does not appear to contribute to the low-level mesocyclone and associated footprint of dynamic lifting. 

\subsubsection{Backward trajectories}

Next we turn to backward trajectories initialized within the most intense upward-moving, cyclonically rotating air in the low-level mesocyclone. While the three layers of tracers show that the low-level mesocyclone is predominately made up of near-ground air, tracers alone cannot show the inflow origins of air bound for the low-level mesocyclone (e.g., the undisturbed, ambient environment versus the forward-flank baroclinic zone). To address this, we initialized backward trajectories within the 90th percentile of vertical velocity at 1 km AGL in each simulation, as this area has the highest potential for stretching of subtornadic surface vortices into tornadoes. 

%\begin{table*}[t]
%\caption{This is a sample table caption and table layout.}\label{t2}
%\begin{center}
%\begin{tabular}{c|ccc}
%\hline\hline
% & \begin{tabular}{c}median trajectory height (m)\\ $+/-$ standard deviation\end{tabular} &  \begin{tabular}{c}90th percentile\\ trajectory height (m)\end{tabular}\\
%\hline
% 0\_0  & \begin{tabular}{c} 190.6 \\ \hspace{6em} $+/-$ 181.5\end{tabular} & 336.5\\
% 1\_0  & \begin{tabular}{c} 562.4 \\ \hspace{6em} $+/-$ 454.1\end{tabular} & 1325.3\\
% 2\_0  & \begin{tabular}{c} 368.7 \\ \hspace{6em} $+/-$ 253.5\end{tabular} & 528.7\\
% 0\_1  & \begin{tabular}{c} 161.9 \\ \hspace{6em} $+/-$ 181.6\end{tabular} & 403.8\\
% 1\_1  & \begin{tabular}{c} 274.0 \\ \hspace{6em} $+/-$ 136.7\end{tabular} & 436.6\\
% 2\_1  & \begin{tabular}{c} 245.1 \\ \hspace{6em} $+/-$ 164.9\end{tabular} & 487.4\\
% 0\_2  & \begin{tabular}{c} 161.4 \\ \hspace{6em} $+/-$ 161.1\end{tabular} & 435.1\\
% 1\_2  & \begin{tabular}{c} 310.0 \\ \hspace{6em} $+/-$ 216.1\end{tabular} & 681.5\\
 %2\_2  & \begin{tabular}{c} 321.0 \\ \hspace{6em} $+/-$ 209.3\end{tabular} & 630.9\\
%\hline
%\end{tabular}
%\end{center}
%\end{table*}

\begin{table*}[t]
\caption{Range of low-level mesocyclone trajectory origin heights for both the median trajectory and the trajectory representing the 90th percentile of data.}\label{t2}
\begin{center}
\begin{tabular}{c|cc}
\hline\hline
 & \begin{tabular}{c}median trajectory\\ origin height (m)\end{tabular} &  \begin{tabular}{c}90th percentile trajectory\\ origin height (m)\end{tabular}\\
\hline
 0\_0  & 190.6 & 336.5\\
 1\_0  & 562.4 & 1325.3\\
 2\_0  & 368.7 & 528.7\\
 0\_1  & 161.9 & 403.8\\
 1\_1  & 274.0 & 436.6\\
 2\_1  & 245.1 & 487.4\\
 0\_2  & 161.4 & 435.1\\
 1\_2  & 310.0 & 681.5\\
 2\_2  & 321.0 & 630.9\\
\hline
\end{tabular}
\end{center}
\end{table*}

\begin{figure*}[t]
\centerline{\includegraphics[width=36pc]{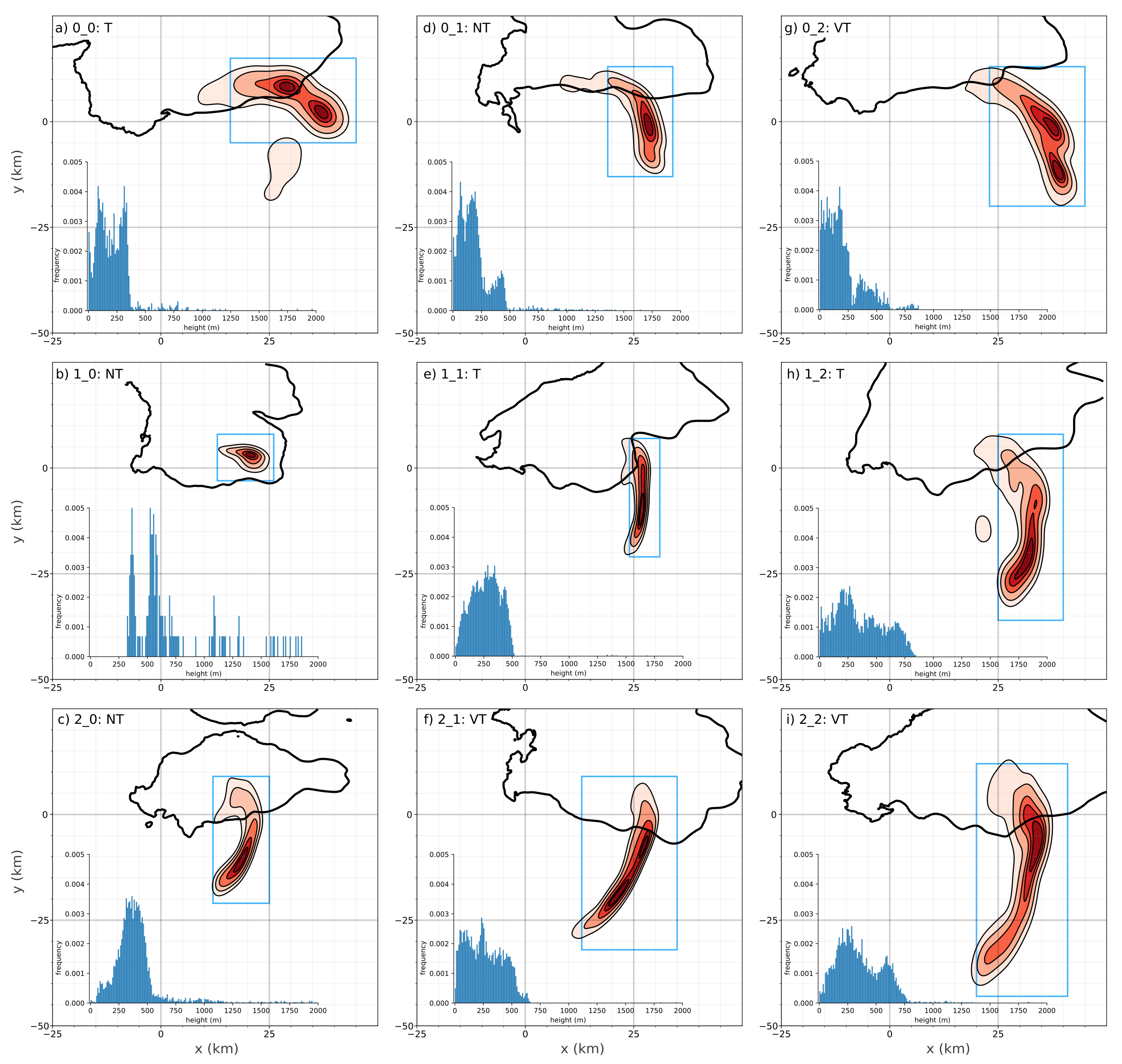}}
\caption{Two-dimensional kernel density estimates (KDE) of the final horizontal locations of the backward trajectories initialized from the low-level mesocyclone after thirty minutes. Shaded KDE values represent 10, 25, 50, 75, and 90\% of the data. The average 10 dBZ reflectivity contour (black) during the five minute composite period ($t_{-10}$ to $t_{-5}$) relative to the key time period of tornado-genesis/failure is shown for the respective simulations. The blue rectangles represent the bounding box used to initialize forward trajectories. Inset in each panel is a histogram of the final vertical locations of the backward trajectories binned every 10 m from 0 to 2000 m AGL. All data is centered upon the 1 km AGL low-level centroid. Each panel is labeled nontornadic (NT), tornadic (T), or violently tornadic (VT).\\} \label{parcel_kde}
\end{figure*}

\begin{figure*}[t]
\centerline{\includegraphics[width=40pc]{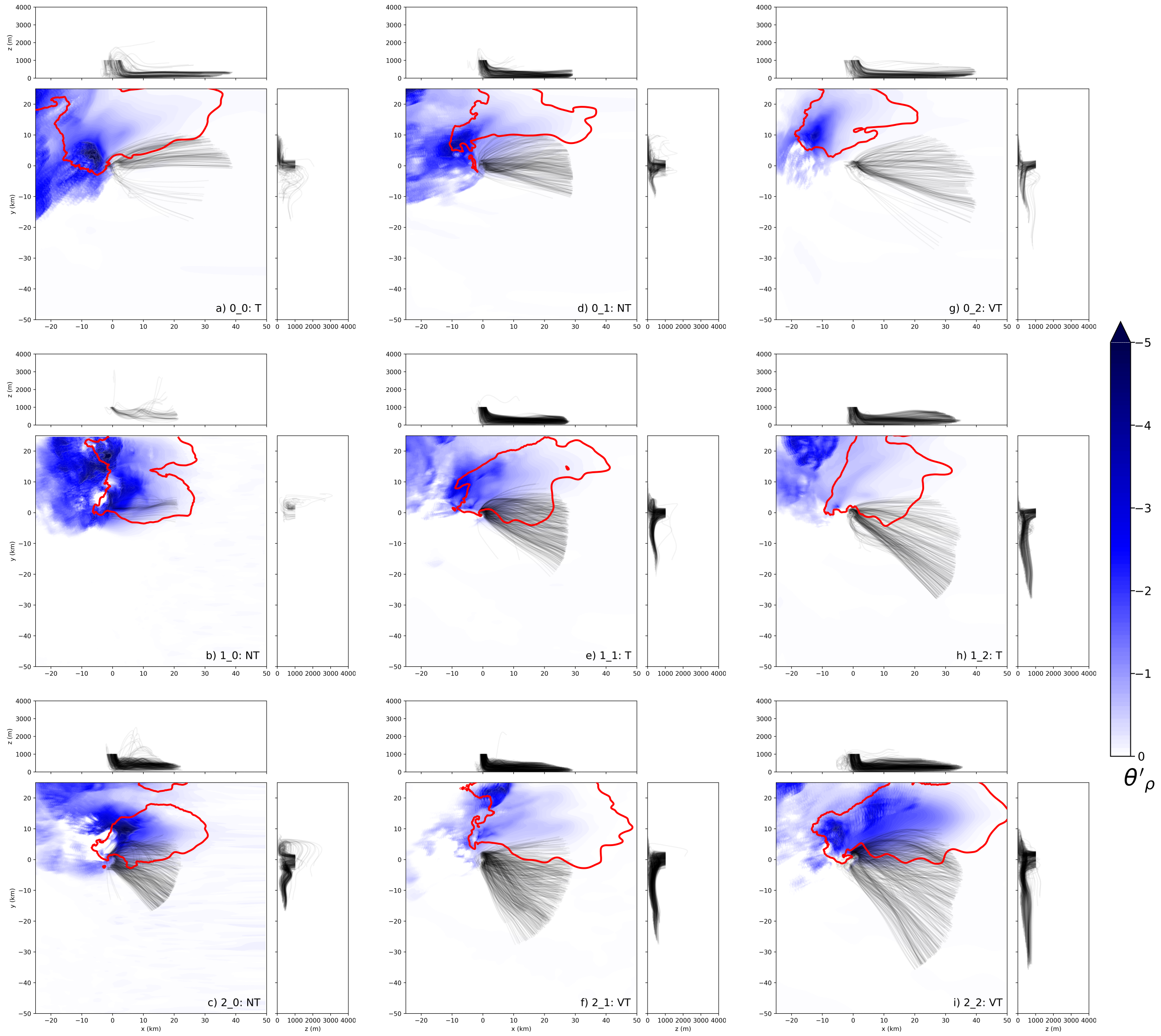}}
\caption{Low-level mesocyclone initialized backward trajectory paths (every 10th trajectory plotted) in the $x-y$, $x-z$, and $y-z$ planes. In the $x-y$ plane, the average density potential temperature perturbation field ($\theta'_{\rho}$) at 10 m AGL and the average 10 dBZ reflectivity contour (red) during the five minute composite period ($t_{-10}$ to $t_{-5}$) relative to the key time period of tornado-genesis/failure is shown for the respective simulations. All data is centered upon the 1 km AGL low-level centroid. Each panel is labeled nontornadic (NT), tornadic (T), or violently tornadic (VT).\\ 
\\} \label{parcel_path}
\end{figure*}

During a five minute period prior to tornado-genesis/failure ($t_{-10}$ to $t_{-5}$), the origin height of inflow air into the mesocyclone is highly consistent across eight of the nine supercells. Similar to the tracer analysis, backward trajectories bound for the low-level mesocyclone originate very close to the ground (Table~\ref{t2}). The distributions of the origin height of trajectories shown in the insets of Figure~\ref{parcel_kde} are generally below 500 m (excluding 1\_0). The median origin height for the eight main supercells is less than 400 m AGL and 90\% of the parcels in each simulation come from below 700 m AGL (Table~\ref{t2}). Only a few trajectories across the matrix of supercells represent ``recycled air'', or air with a history of descent from farther aloft\footnote{The exact path of ``recycled'' low-level mesocyclone trajectories should be treated with caution since the likelihood of errors in the backward trajectory integration is higher for such a flow regime. Regardless, the overwhelming proportion of trajectories that rise into the low-level mesocyclone from the undisturbed inflow compared to the ``recycled air'' is still qualitatively informative.} (Fig.~\ref{parcel_path}). Many of the simulations have median parcel heights less than 300 m (Table~\ref{t2}). The very low altitude of parcels that contribute to the strongest vertical motion within the mesocyclone likely explains the comparative forecast skill of environmental streamwise horizontal vorticity and thus environmental SRH in progressively shallower layers \citep[e.g., as shallow as 0 -- 250 m AGL in][]{coffer2020era5}.

Compared to the vertical extent of the inflow, the horizontal extent of air bound for the low-level mesocyclones across the simulations generally originates from south and east of the low-level updraft (Figs.~\ref{parcel_kde},\ref{parcel_path}), consistent with the orientation of the near-ground hodographs in Figure 3. The trajectory fields in the simulations with higher SRH have a noticeably more expansive inflow region, especially towards the southeast (Fig.~\ref{parcel_kde}e-i). The highest density of parcel origins in most of the simulations (Fig.~\ref{parcel_kde}d-i) is from well outside of the precipitation field. The paths of these trajectories, coming from the undisturbed, far-field environment into the low-level mesocyclone, appear to traverse the forward flank only minimally (or not at all in some instances, e.g., Fig.~\ref{parcel_path}d,g). Especially for the tornadic supercells, the parcel origins are mostly from the far-field, toward the southeast (Fig.~\ref{parcel_kde}e-f-i), with one exception (0\_0). In 0\_0, the initial locations are primarily due east of the low-level mesocyclone (Fig.~\ref{parcel_kde}a) and flow parallel to the forward flank (Fig.~\ref{parcel_path}a), consistent with the orientation of the storm-relative wind in the 0\_0 hodograph (Fig.~\ref{hodographs}a). The prevalence of parcels originating from the undisturbed inflow environment is consistent across multiple possible definitions of a ``low-level mesocyclone''. Coherent areas of large, positive circulation at 1 km AGL display very similar  trajectory origins and statistics (see Supplemental Figs. 1-3), due to a high degree of correlation between the areas of large circulation and large vertical velocity (not shown).

\begin{figure*}[t]
\centerline{\includegraphics[width=40pc]{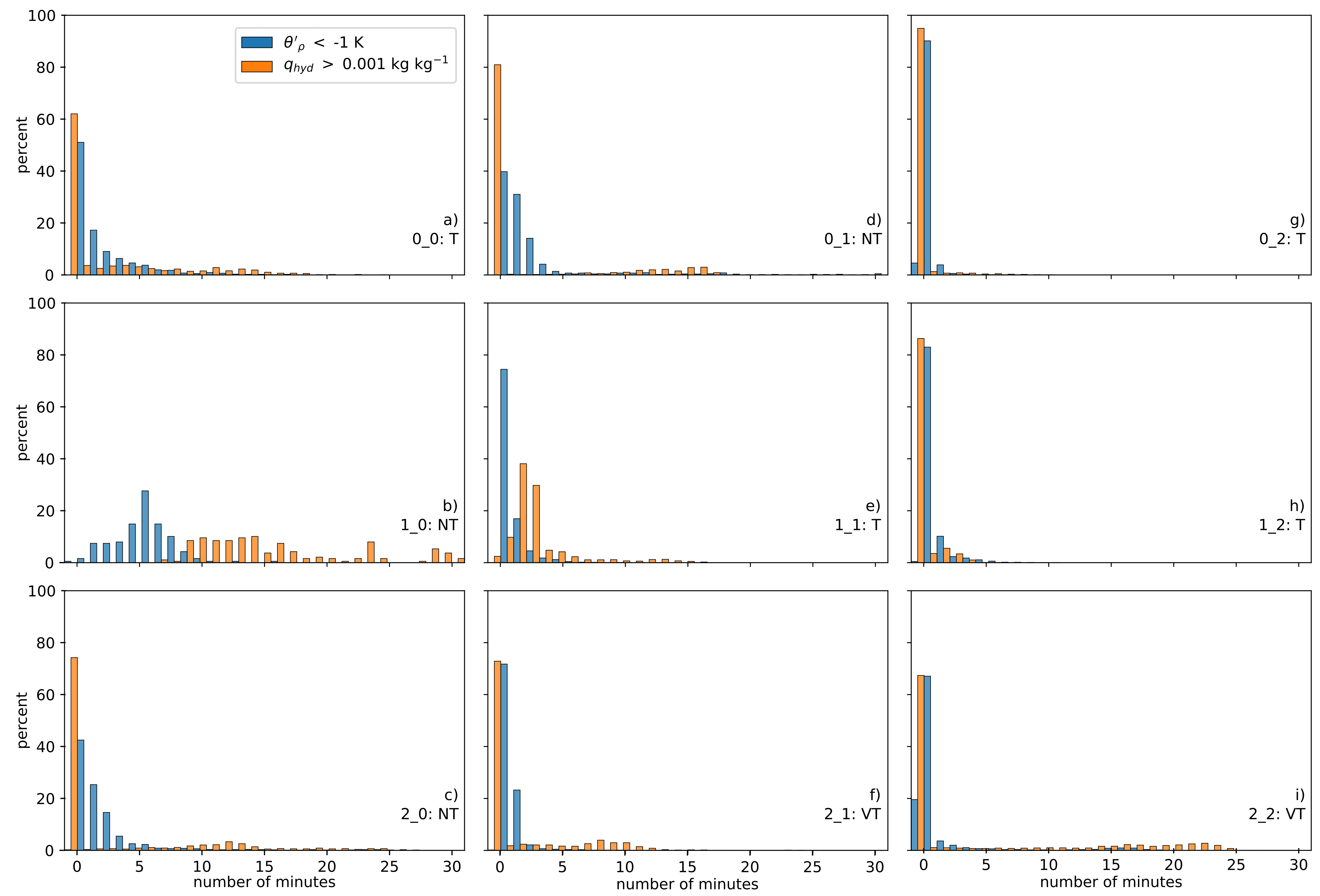}}
\caption{Histograms of the accumulated time each backward trajectory initialized from low-level mesocyclone spent in regions characterized by $\theta'_{\rho} <$ -1 K (blue) and $q_{hyd}$ > 0.001 kg kg$^{-1}$ (orange) for each simulation. Data is binned in 1 minute increments. The area underneath each histogram sum to 100\%.  Each panel is labeled nontornadic (NT), tornadic (T), or violently tornadic (VT).} \label{parcel_histograms}
\end{figure*}

At least 65\% of parcels bound for the low-level mesocyclone in the eight main supercells spend less than 5 minutes in areas influenced by the storm's hydrometeor and negative buoyancy fields (Fig.~\ref{parcel_histograms}), based on the accumulated time within regions characterized by $q_{hyd} > 0.001$ kg kg$^{-1}$ and $\theta'_{\rho} < -1$ K. These two, admittedly arbitrary, thresholds only provide an estimate of the time the backward trajectories spent within 'storm outflow', not the potential for baroclinic streamwise vorticity generation (which is explored more thoroughly using the material stencils in the subsequent subsection). In the instances characterized by the weakest forward flank cold pools (Fig.~\ref{parcel_path}f,g,h), this percentage is greater than 90\%. For the small percentage of parcels that do interact with baroclinic gradients associated with the forward flank, the experienced deficits in density potential temperature rarely exceed -2 to -3 K, and are generally closer to -1 K (Fig.~\ref{parcel_path}). Cold pool deficits are often even weaker 30 minutes prior to tornado-genesis/failure, when the trajectories were initialized, than Figure~\ref{parcel_path} would suggest (not shown). In that respect, the density potential temperature fields in the present simulations resemble those of the observed tornadic supercells in \citet{shabbott2006surface}. As a result of the short residence time within the storms' forward flanks, weak deficits in potential temperature, and fast storm-relative winds accelerating towards the supercell, the mean low-level mesocyclone trajectory in the eight main supercells experiences a rather small change in streamwise horizontal vorticity along its inflow path [estimated using Eq. 1 from \citet{shabbott2006surface}] . This is more precisely quantified with the material stencils in the following section. 
%at least an order of magnitude less than the respective base-state environments (not shown), based on a estimate from the scale analysis presented by \citet{klemp1983study} and \citet[, see their Eq. 1]{shabbott2006surface}. 

In summary, inflow air into the low-level mesocyclone originates very close to the ground and overwhelmingly from the undisturbed, far-field environment (toward the southeast). Most parcels bound for the low-level mesocyclone experience minimal effects from the storm's precipitation field. Both of these results would be expected to have a direct effect on the importance of environmental versus storm-generated vorticity contributions to the low-level mesocyclone. We explore this topic directly next.

\subsection{Contributions of environmental vs. storm-generated vorticity to the low-level mesocyclone}

Through both tracers and backward trajectories, we have shown thus far that the air comprising the most intense upward-moving, cyclonically rotating air in the low-level mesocyclone originates from the near-ground layer ($<$500 m AGL) and predominately from the undisturbed inflow environment, with the highest density of parcels appearing to have very little residence time within the region of precipitation and negative buoyancy associated with the forward-flank region. On the face of it, these two factors would seemingly implicate the near-ground environmental horizontal streamwise vorticity as the dominant contributor to the overall rotation of the low-level mesocyclone, not the storm-generated streamwise vorticity classically associated with low-level mesocyclone-genesis. To quantitatively evaluate this interpretation, we track forward trajectories bound for the low-level mesocyclone and assess their associated vorticity via stencils of nearby adjacent parcels following the technique described by \citet{dahl2014imported}. As described in Section~\ref{methods}, forward trajectories were seeded within model restart files upstream of the low-level mesocyclone within a unique horizontal bounding box for each simulation (the blue boxes in Fig.~\ref{parcel_kde}) encompassing at least 75\% of the low-level mesocyclone inflow area based on the origins of the backward trajectories. The seeding of the forward trajectories is meant to represent a majority of the inflow air; computational limitations prevent an exhaustive sample of all possible inflow air. %Nevertheless, based on the other analysis techniques presented above, this setup is still suitable for the purpose of determining the dominant contribute to the low-level mesocyclone's rotation.

\begin{figure*}[t]
\centerline{\includegraphics[width=40pc]{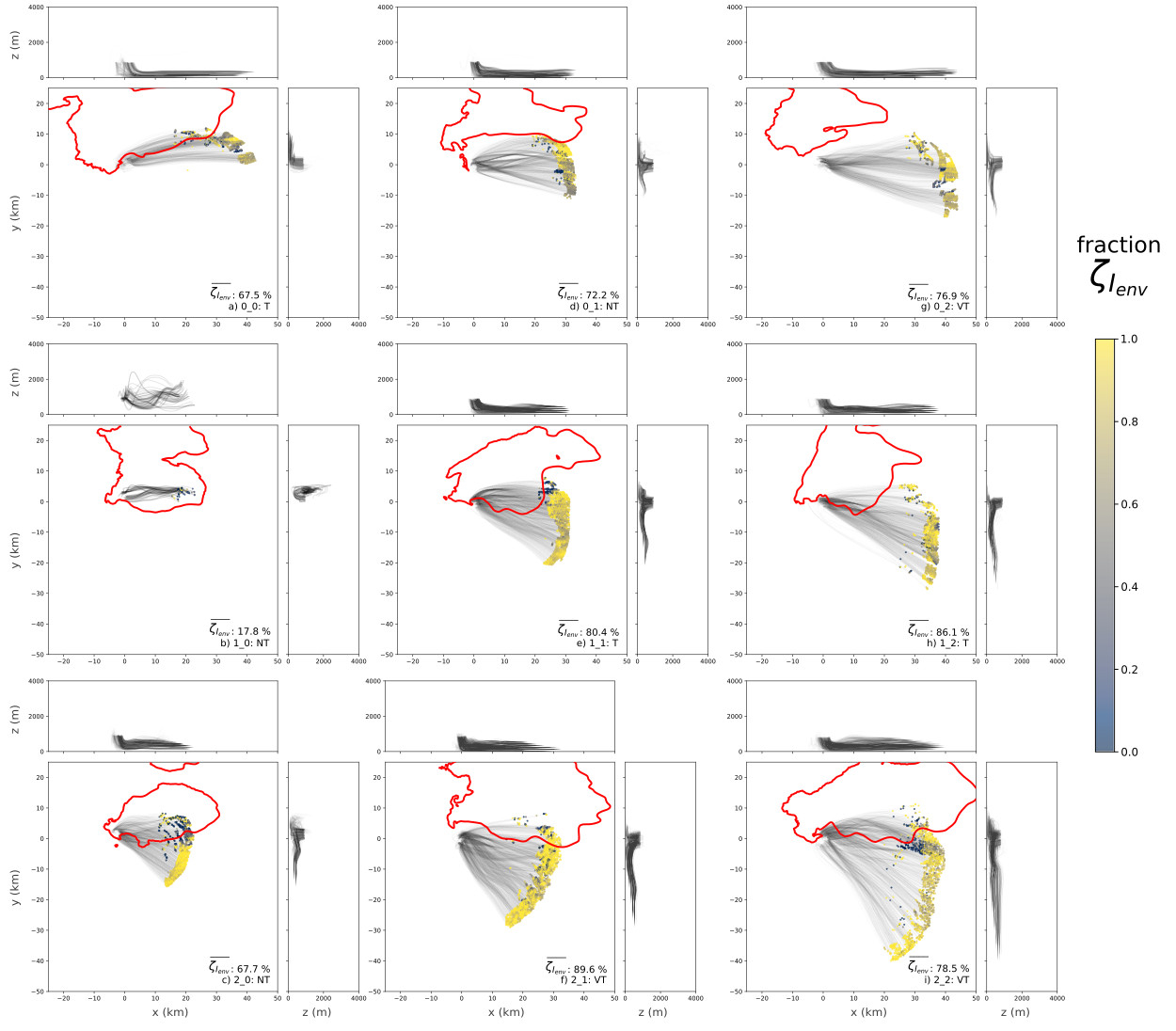}}
\caption{Paths of forward trajectories that end up in the low-level mesocyclone in the $x-y$, $x-z$, and $y-z$ planes. Also shown in the $x-y$ plane, a scatter plot of the initial location of each forward trajectory colored by the fraction of $\zeta_{I_{env}}$ / $\zeta_{LLM}$ from 0 to 1 and the average 10 dBZ reflectivity contour (red) during the five minute composite period ($t_{-10}$ to $t_{-5}$) relative to the key time period of tornado-genesis/failure is shown for the respective simulations. All data is centered upon the 1 km AGL low-level centroid. Each panel is labeled nontornadic (NT), tornadic (T), or violently tornadic (VT) and shows the median $\overline{\zeta_{I_{env}}}$ expressed as a percentage from Figure~\ref{stencils}. 
\\} \label{parcel_BTpercent}
\end{figure*}

The forward trajectory parcels that meet the vertical velocity and vertical vorticity thresholds of the low-level mesocyclone [vertical velocities greater or equal to the 90th percentile at 1 km AGL (Table~\ref{t1}) and at least 0.01 s$^{-1}$ of vertical vorticity (within $+/-$ 10 m of 1 km AGL)]\footnote{An overwhelmingly majority of the forward trajectories released within the inflow region would have qualified as low-level mesocyclone parcels if the parcel output frequency was decreased and/or the depth of the vertical layer surrounding 1 km AGL was increased. These choices simply acted to filter parcels to a reasonable number for analysis given storage and computational constraints.} have similar paths into the low-level mesocyclone and originate from similar locations as the backward trajectories in the previous section (Fig.~\ref{parcel_BTpercent}). This provides some quality assurance, which is welcome in light of documented differences in accuracy between forward and backward trajectory techniques \citep{dahl2012uncertainties}. Because the forward trajectories were not seeded at the exact terminal locations of the backward trajectories (rather, they were initialized over an isotropic grid covering most of the inflow region), it is not possible to directly compare the two sets of trajectories \citep[as-in][Gowan et al. 2021]{dahl2012uncertainties}; however, many of the details, including the shape, width, depth, and proportion of undisturbed, far-field environment parcels to forward flank parcels, are extremely similar (Figs.~\ref{parcel_path},~\ref{parcel_BTpercent}).%, despite the aforementioned built-in memory of the backward trajectories.  

\begin{figure*}[t]
\centerline{\includegraphics[width=40pc]{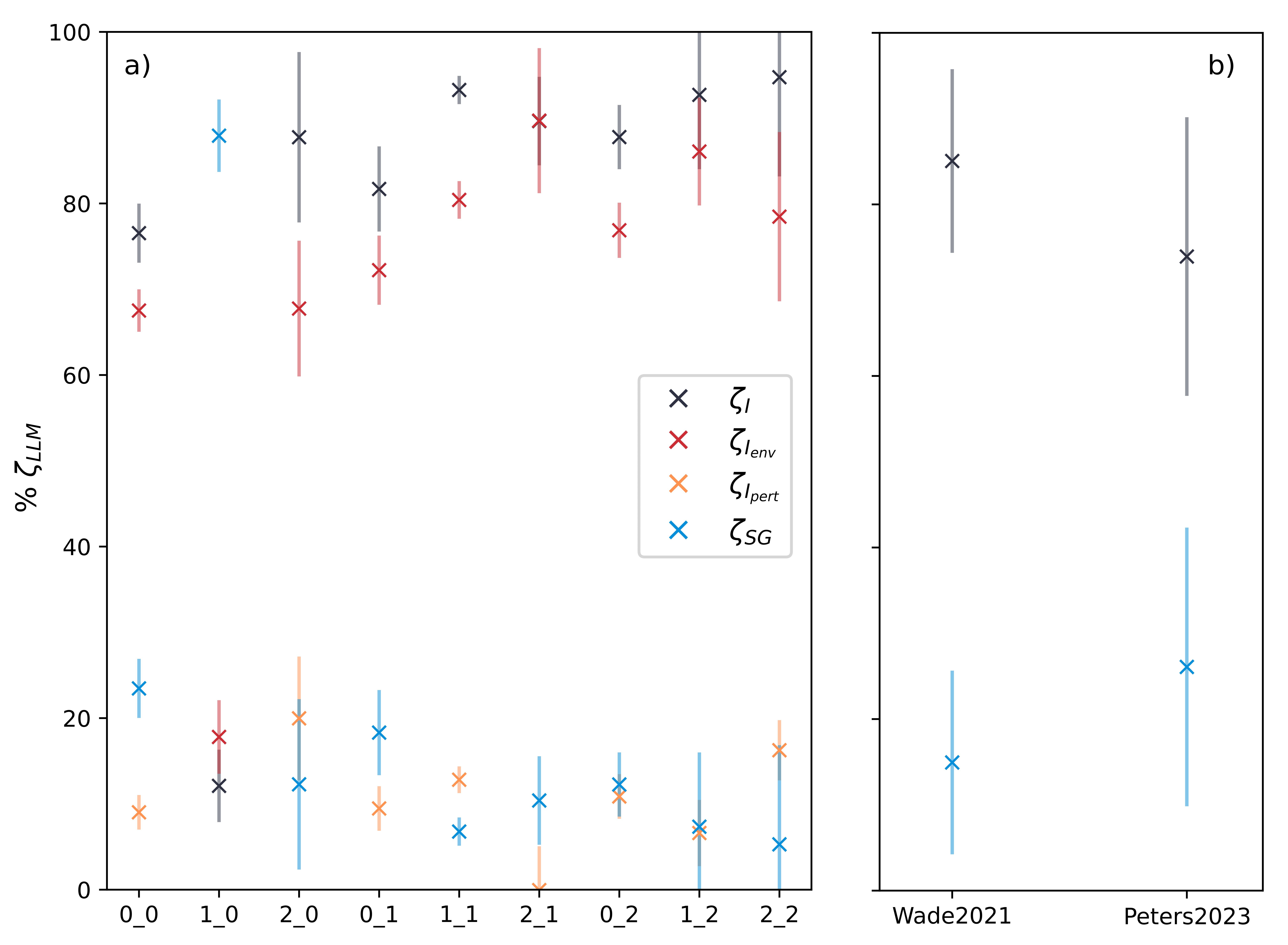}}
\caption{Percentage of $\zeta_{LLM}$ attributable to $\zeta_{I}$, $\zeta_{I_{env}}$, $\zeta_{I_{pert}}$, and $\zeta_{SG}$ for low-level mesocyclone parcels from both a) the supercell simulations presented herein and b) from the supercell simulations by \citet{wade2021dynamics} and \citet{peters2023disentangling}. In a), `x' marks the median percentage value from all the low-level mesocyclone trajectories for each simulation and the error bars represent the standard deviation of those trajectories, while in b), `x' marks the median percentage from the average  trajectory for each simulation [\citet{wade2021dynamics}: n=4, \citet{peters2023disentangling}: n=24] and the error bars represent the standard deviation of each paper's simulations.}\label{stencils}
\end{figure*}

Each seven-parcel stencil tracks $\zeta_{I_{env}}$ and $\zeta_{I_{pert}}$ (as described in Section~\ref{methods}) via the deformation and stretching of the initial (or ``imported'') vorticity vector over time, while $\zeta_{SG}$ (and the rearrangement of $\zeta_{SG}$) is computed as a residual. In eight of the nine simulations (except 1\_0),  $\zeta_{I}$ is far and away the dominant contributor to the low-level mesocyclone vertical vorticity\footnote{The percentage of $\zeta_{I}$ and $\zeta_{SG}$ to the total $\zeta_{LLM}$ was calculated via both a simple ratio as well as a weighted average by the magnitude of $\zeta_{LLM}$. Percentages from either method resulted in very similar results presented in Figs.~\ref{parcel_BTpercent},\ref{stencils}; however, in general, the contribution from $\zeta_{SG}$ using the weighted average was approximately 3-5\% higher across the simulations, implying parcels that develop additional vorticity from storm-generated sources contribute slightly more to strongest rotation of the low-level mesocyclone.} ($\zeta_{LLM}$; Fig.~\ref{stencils}a). Only in simulation 1\_0, the weakest and most transient supercell, is $\zeta_{SG}$ the dominant component of $\zeta_{LLM}$. This is not entirely unexpected considering the entire inflow region of the 1\_0 supercell's low-level mesocyclone originates within precipitation of the forward flank (Figs.~\ref{parcel_kde}b,~\ref{parcel_path}b,~\ref{parcel_BTpercent}b). For the other eight main simulations, $\zeta_{I_{env}}$ contributes between 65\% and 90\% of the total $\zeta_{LLM}$ indicating that \emph{the environmentally-derived vorticity comprises a much larger percentage of the mesocyclone's vertical vorticity than the storm-generated vorticity} (Fig.~\ref{stencils}a). The dominant contribution of $\zeta_{I_{env}}$ to the total $\zeta_{LLM}$ is consistent across multiple possible definitions of a ``low-level mesocyclone'', including for lower altitudes than 1 km AGL (specifically at 750 and 500 m AGL; see Supplemental Fig. 4) and for a coherent area of positive circulation (regardless of vertical velocity and vertical vorticity values; see Supplemental Figs. 1-3,5).

%Considering the path of the parcel trajectories bound for the low-level mesocyclone (Fig.~\ref{parcel_BTpercent}), we can surmise that the regions of enhanced streamwise vorticity (Fig.~\ref{SVC}) in each of these supercells is primarily due to horizontal stretching of environmental vorticity as near-ground air accelerates towards the low-level updraft, similar to the vorticity budgets of \citet{schueth2021svc}. This supports the argument that stronger low-level updrafts yield more intense storm-enhanced regions of streamwise vorticity (not vice versa). 

While $\zeta_{I_{env}}$ is generally quite high for all these eight simulations, there is a trend for the tornadic supercells towards the right of Figure~\ref{stencils}a, starting with 1\_1, to have a higher percentage of $\zeta_{I_{env}}$ than the nontornadic supercells. These tornadic simulations also have the most favorable lower tropospheric base-state hodographs and the highest values of SRH500. The exception to that trend, simulation 0\_0, has $\sim$20\% lower contribution from $\zeta_{I_{env}}$ than the other tornadic supercells (Fig.~\ref{stencils}a). Not only does the 0\_0 supercell have a base-state SRH500 of 67 $m^{2} s^{-2}$ \citep[well below the median tornadic SRH500 value from][]{coffer2019srh500}, but also has the lowest concentration of near-ground tracer among the tornadic low-level mesocyclones (Fig.~\ref{tracer_percent}) and highest proportion of parcels that flow into the mesocyclone directly parallel to the forward flank baroclinic gradient (Fig.~\ref{parcel_path}). In total, this potentially suggests the 0\_0 supercell required additional augmentation from within storm baroclinic generation of streamwise vorticity to establish a low-level mesocyclone capable of producing a tornado. Fully fleshing out this hypothesis would require an ensemble of simulations and more additional analysis, which is beyond the scope of this study and will be expanded upon in future work.  

The next largest contributor to $\zeta_{LLM}$ (either $\zeta_{I_{pert}}$ or $\zeta_{SG}$) varies between the individual simulations but is generally less than 20\%. There is no discernible trend for the nontornadic or tornadic supercell simulations to have more or less $\zeta_{I_{pert}}$ than $\zeta_{SG}$. As a reminder, because we cannot say whether the $\zeta_{I_{pert}}$ represents prior reorientation or stretching of base-state vorticity versus prior baroclinic (or frictional) generation, it is treated separately. Even if we generously assume $\zeta_{I_{pert}}$ is entirely attributed to storm-generated effects, their combined contribution would still be less than 35\% of the total $\zeta_{LLM}$ for the eight main storms. 

%It has usually been unclear in prior studies how much prior baroclinic generation has already occurred \citep{dahl2014imported}, which then affects the final attribution of $BT$ versus $NBT$. We hope future studies will also parse out $BT_{env}$ from $BT_{pert}$ to help alleviate those criticims. 

While the bulk percentages of $\zeta_{I}$ compared to $\zeta_{SG}$ paint a clear picture that most of the low-level mesocyclone rotation is from the environmental vorticity, examining individual parcels and their ratio of $\zeta_{I_{env}}$ to $\zeta_{LLM}$ highlights source regions from which generation from the storm is more prominent, such as the forward flank. There is a trend in some (but not all) of the supercells to have lower percentage of $\zeta_{I_{env}}$ within parcels that originate closer to the hydrometeor field and forward flank (Fig.~\ref{parcel_BTpercent}c,e,i). Those three simulations (2\_0, 1\_1, 2\_2) also represent simulations where the backward trajectories cross through larger density potential temperature gradients (Fig.~\ref{parcel_path}c,e,i) and have a higher frequency of parcels that spend 5 -- 25 minutes of accumulated time within the hydrometeor and negative buoyancy fields (Fig.~\ref{parcel_histograms}c,e,i). Specifically looking at simulation 2\_2 (Fig.~\ref{parcel_BTpercent}i) as an example, many of the parcels with the lowest percentage of $\zeta_{I_{env}}$ (and thus highest $\zeta_{SG}$; located at approximately x=30, y=-5 in Fig.~\ref{parcel_BTpercent}i) traverse west-northwestward into forward flank (and larger negative buoyancy gradients; Fig.~\ref{parcel_path}i), before turning back towards the updraft and eventually rising into the low-level mesocyclone. Due to their path through the storm, and exposure to horizontal baroclinity, these isolated parcels likely correspond to the classical conceptual model of forward flank air being re-ingested into the low-level updraft and wall cloud \citep[i.e., ][see their Fig. 5]{atkins2014observations}. %This suggests in supercells with larger buoyancy deficits in the forward flank, $\zeta_{NBT}$ might contribute more prominently \citep[although larger are typically associated with nontornadic supercells][]{shabbott2006surface}

\begin{figure*}[t]
\centerline{\includegraphics[width=22pc]{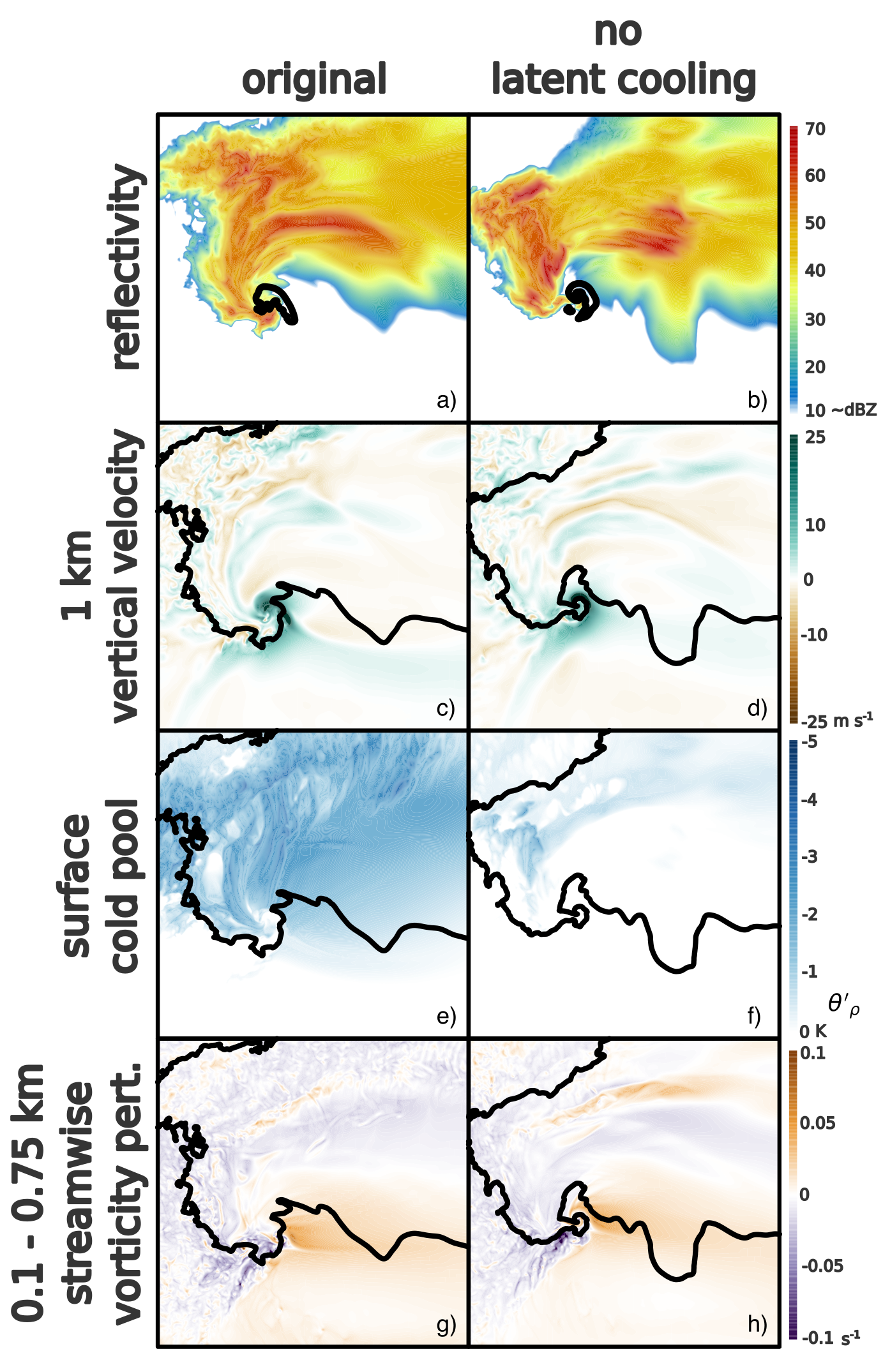}}
\caption{Comparison of the original 2\_2 simulation (left) and a 30 minute rerun without evaporation, melting, or sublimation (i.e., “no latent cooling”; right) at the key time period of tornadogenesis (t = 143 min) for (top row) 10 m AGL reflectivity (dBZ; shaded) and the 90th percentile of updraft at 1 km (black contour), (second row) 1 km AGL vertical velocity (m s$^{-1}$; shaded) and 10 dBZ reflectivity (black contour), (third row) 10 m AGL density potential temperature perturbation (K, shaded) and 10 dBZ reflectivity (black contour), (bottom row) 0.1 - 0.75 km AGL average streamwise vorticity perturbation (s$^{-1}$; shaded) and 10 dBZ reflectivity (black contour).} \label{noLC}
\end{figure*}

As a final test of the apparent unimportance of baroclinic generation within the forward flank to the low-level mesocyclone, the 2\_2 simulation was rerun without evaporation, melting, or sublimation (i.e., no latent cooling) for the 30 minutes prior to the key time period of tornadogenesis. \emph{Despite having almost no remaining cold pool, a similarly intense low-level mesocyclone occurs} in the cooling-free experiment ($w_{1KM}$ > 40 m s$^{-1}$; Fig.~\ref{noLC}c,d). And, an almost-identical region of enhanced streamwise vorticity within the near-inflow region also persists when microphysical cooling is turned off ($w_{1KM}$ > 40 m s$^{-1}$; Fig.~\ref{noLC}g,h). Although we cannot rule out the possibility of some prior memory of a baroclinic forcing and/or a convergent boundary within the forward-flank, there is essentially no lasting negative buoyancy present within the forward flank (Fig.~\ref{noLC}f; besides the minimal contribution from hydrometer drag). Thus, the SVC-like feature present must be due to horizontal stretching of the ambient, environmental vorticity (contained within the term $\zeta_{I_{env}}$ in the realm of material stencils) in response to the inflow low associated with the updraft. \citet{schueth2021svc} also found that the maximum vorticity in their simulated SVC was almost solely driven by horizontal stretching. This sensitivity test underscores the importance of the environmental vorticity since a similarly-intense low-level mesocyclone occurs in the absence of baroclinic generation.  %Cross sections through this region resemble the SVC structure associated with a Kelvin-Helmholtz billow in \citet[see their Fig. 8]{schueth2021svc}. 

\subsubsection{Low-level mesocyclone vorticity sources in supplemental simulations}

On one hand, because of the documented forecast skill of near-ground environmental SRH in separating nontornadic from significantly tornadic supercell environments, it seems entirely logical that the environmental vorticity would exert a substantial influence on the total vertical vorticity of the low-level mesocyclone. On the other hand, because the low-level mesocyclone has often been attributed to baroclinically-generated streamwise horizontal vorticity within the storm (as discussed in the introduction), the degree to which $\zeta_{I_{env}}$ dominates, and consistency among the simulations, is somewhat surprising. To supplement the results from the simulations presented herein, we also present previously unpublished material stencil analyses from two existing supercell studies in the literature, \citet{wade2021dynamics} and \citet{peters2023disentangling}.

These studies complement our simulations by virtue of their different thermodynamic and kinematic environmental profiles. \citet{wade2021dynamics} focused on three high-shear, low-CAPE (HSLC) environments from a VORTEX-SE case in Alabama on 31 March 2016 as well as a companion high-shear, high-CAPE environment from the 3 April 1974 ``Super Outbreak'' (see their Fig. 6 for Skew $T$-$logP$ and hodograph diagrams).  \citet{peters2023disentangling} presented a large number of supercell simulations with ``L'' and ``C'' shaped hodographs (see their Figs. 1-3 for Skew $T$-$logP$ and hodograph diagrams), independently varying the streamwise vorticity and storm-relative flow to disentangle their influence on low-level mesocyclone characteristic. Each simulation in \citet{peters2023disentangling} used a constant thermodynamic environment based on the tornadic VORTEX2 composite environment from \citet{parker2014composite}. The reader is referred to these papers for more details about their simulations. The material stencil analysis was conducted independently amongst the three studies, with varying thresholds of vertical velocity/vertical vorticity and criteria for when/where forward trajectories were seeded. Of note, these studies calculated $\zeta_{I}$ and $\zeta_{SG}$ only; no attempt was made to distinguish $\zeta_{I_{env}}$ from $\zeta_{I_{pert}}$. The chief similarity between all three studies is that parcels were filtered to highlight those entering the low-level mesocyclone at 1 km AGL. We believe the modest differences in analysis techniques increase confidence that the results presented in the previous subsection are not unique to our specific methodological choices. 

Similar to the eight main supercell simulations in the present study, the low-level mesocyclones of both \citet{wade2021dynamics} and \citet{peters2023disentangling} primarily derive vertical vorticity from the environment rather than storm itself (Fig.~\ref{stencils}b). In the \citet{wade2021dynamics} storms, the low-level mesocyclone is almost entirely environmentally driven for the parcel groups in the high-CAPE storm and low-CAPE 1 and 3 ($>$ 85\%). Even in their low-CAPE 2 supercell, which in general presented more analysis challenges than the other three supercells [see \citet{wade2020phd} for more details], the initial environmental component of the low-level mesocyclone is greater than 65\%. Given the differences in storm structures and cold pools amongst the high-CAPE and low-CAPE storms \citep{wade2021dynamics}, it is remarkable how consistently little the storm-generated term contributes to the low-level mesocyclone (Fig.~\ref{stencils}b, Supplemental Fig. 6). For the \citet{peters2023disentangling} supercells, the storm-generated component comprised less than 35\% of the total $\zeta_{LLM}$ on trajectories in most simulations (Fig.~\ref{stencils}b), only exceeding this percentage in the storms with the weakest mesocyclones. Regardless of the differing combinations of storm-relative flow and streamwise vorticity, the environmental contribution was generally greater than 65\%\footnote{None of the low-level mesocyclone/updraft attributes systemically evaluated in \citet{peters2023disentangling}, such as updraft and mesocyclone radius, net updraft circulation and rotational velocity, as well as average updraft vertical vorticity and helicity density, displayed any meaningful correlation with the fraction of $\zeta_{I}$ to $\zeta_{LLM}$ (Supplemental Fig. 7).}. In total, these results further demonstrate that \emph{environmentally-derived vorticity comprises a much larger percentage of the low-level mesocyclone than storm-generated vorticity in persistent, mature supercells}. 

\section{Conclusions and discussion}

In this article, we sought to address where inflow air bound for the low-level mesocyclone originates from and whether the origins of such air could address the dynamical role of near-ground streamwise vorticity present in the ambient environment versus what is generated in-situ within the forward flank of a supercell. The streamwise vorticity present within the environment, and the augmentation of the vorticity by the storm, can potentially modulate the intensity the low-level mesocyclone and ultimately determine whether a supercell fails or succeeds at producing a tornado. Using a matrix of nine supercell simulations, initialized with a spectrum of near-ground wind profiles observed in nature, we found the following: 

\begin{itemize}
    \item The air that comprises the core of the mesocyclone at 1 km, where the greatest potential for vertical stretching exists, originates almost exclusively from very close to the ground, often in the lowest 200 - 400 m AGL. Air originating above 500 m AGL does not tend to contribute to the main updraft until farther aloft. 
    \item Air bound for the low-level mesocyclone primarily originates from the undisturbed, ambient environment, rather than from along the forward flank. In both the nontornadic and tornadic supercells, 60 to 90\% of the inflow air into the low-level mesocyclone has little to no residence time within regions characterized by negative buoyancy and hydrometeors in the forward flank. 
    \item The dominant contributor to vertical vorticity within the low-level mesocyclone is from the environmental horizontal vorticity, with storm-generated providing very little augmentation to the low-level mesocyclone. As much as 90\% of the low-level mesocyclone vertical vorticity can be solely attributed to the base-state environment. For the few parcels that do traverse baroclinic gradients within the storm, the augmentation to the low-level mesocyclone from storm-generated vorticity is higher. 
\end{itemize}

We were motivated to understand why near-ground environmental streamwise vorticity is such a highly skillful tornado forecast parameter given the long history linking the low-level mesocyclone intensity to within-storm baroclinic generation of horizontal vorticity \citep[e.g.,][]{klemp1983study,rotunno1985rotation}. Low-level mesocyclone air in our simulated supercells comes from the ambient environment. Thus, low-level mesocyclone does not require much augmentation from the development of additional horizontal vorticity in the forward flank. %In some ways, in the classical conceptual model of low-level mesocyclone-genesis, we want all the benefits of a robust forward flank baroclinic gradient, without any of the detrimental effects facts that come along with negatively buoyant outflow. 
The ingestion of forward flank parcels may instead be a \emph{symptom} of an intensifying low-level mesocyclone (driven by environmental horizontal vorticity). After all, in order for rain-cooled air to be re-ingested by the storm, substantial dynamic lifting must be present to force negatively (or at least neutrally) buoyant forward flank outflow parcels upwards. 

Perhaps some of the apparent prior ambiguity within the past literature can simply be attributed to differences in nomenclature between studies (or lower vertical resolution). Many of the earlier references to 'low-level mesocyclones' within the literature discussed rotation approximately within the lowest 250 m AGL \citep[e.g.,][]{rotunno1985rotation,brooks1993environmental,brooks1994role,wicker1995simulation,gilmore1998influence,adlerman1999numerical,atkins1999influence,ziegler2001evolution,beck2013assessment}. Over time, with increased emphasis on the supercell's footprint of dynamic lifting being established at relatively low altitudes, about 1 km above the ground, \citep[e.g.,][]{markowski2014influence}, a distinction between ``near-ground rotation'' and ``low-level mesocyclone rotation'' has developed in the literature. These earlier studies were likely quantifying the development of \emph{near-ground} rotation, which has rather conclusively been attributed to baroclinic gradients and vorticity generation within \emph{downdrafts} in the outflow \citep[summarized in][]{fischer2022transition}. On the other hand, the low-level mesocyclone is predominately associated with \emph{upward} tilting of horizontal vorticity\footnote{Even if the forward flank contributes more to the overall rotation of the low-level mesocyclone than shown herein, these highly streamwise parcels in the lower troposphere would be tilted into the mesocyclone by an updraft, not a downdraft.}. Despite the apparent evolving definition of term 'low-level mesocyclone', there is still widespread reference to \citet{rotunno1985rotation} in recent years regarding the development of low-level mesocyclone rotation, even when clearly discussing rotation near the cloud base \citep[e.g.,][]{orf2017evolution,frank2018entropy, markowski2020intrinsic,fischer2020relative,flournoy2020modes,flournoy2021motion,murdzek2020processes,murdzek2020svc,schueth2021svc,davies2022theory,finley2023svc}. We encourage future research endeavours to use a unified nomenclature when discussing supercell processes in the lower troposphere, as the governing dynamics between the development of rotation near the ground ($<$ 250 m AGL) and near cloud-base ($\sim$ 1 km AGL) are very likely distinct (at least prior to tornadogenesis). 

Future studies could also clarify the conceptual distinction between ``low-level'' and ``mid-level'' mesocyclones. If both are generated via the upward tilting of environmental horizontal vorticity, at what point does the lowest portion of the mid-level mesocyclone impact the rotation near cloud-base? Are mesocyclones truly bi-modal as the community has often defined them? Vertical profiles of vertical vorticity leading up to the tornadic phase of a supercell in \citet[their Fig. 6]{klemp1983study}, as well as across the matrix of simulations presented herein (Supplemental Fig. 8), appear to substantiate the distinction of the low-level mesocyclone near cloud-base from the mid-level mesocyclone. Across all nine simulations, there is a specific local maximum in the vertical vorticity field at approximately 1 km AGL that potentially provides the essential dynamic lifting below the LFC (where buoyancy cannot make a positive contribution) needed for tornadogenesis. In total, this suggests that the low-level mesocyclone is not simply the bottom contour of the mid-level mesocyclone; however, a more thorough analysis is warranted across a wider range of supercell environments and storm structures. 

The operational ramifications of this study include the explanation that SRH500 is a more skillful tornado parameter than SRH over deeper layers because air near the ground is overwhelmingly more likely to contribute to the low-level mesocyclone than air farther aloft. In turn, horizontal vorticity of the near-storm inflow exerts substantial control over the width and intensity of the low-level mesocyclone \citep{peters2023disentangling}. These results further underscore the need for more frequent and more numerous observations of the near-ground vertical wind profile than what are currently available. Improvements to forecasters' situational awareness could be realized through high spatial and temporal sampling of the environment near storms \citep[e.g., as shown by][]{chilson2019moving,bell2020confronting}. 

Our simulations also provide three-dimensional context for the observations of forward-flank outflows presented by \citet{shabbott2006surface}, who found tornadic supercells to have small density gradients within the forward flank, whereas streamwise vorticity generation was largest in nontornadic cases. As speculated by \citet{shabbott2006surface}, ``large ambient streamwise vorticity might obviate the need for baroclinic streawmwise vorticity production'', and in fact, ``substantial baroclinic vorticity generation in the [forward-flank] outflow might be unfavorable for low-level mesocyclones and tornadogenesis" because excessive negative buoyancy, whether in the rear-flank or the forward-flank is generally unfavorable for tornadoes \citep{markowski2002direct,grzych2007thermodynamic,hirth2008surface,weiss2015comparison,bartos2022balloon}. This seems to be borne out by our simulations. Because of the out-sized role of the ambient environment in low-level mesocyclone-genesis herein (regardless of the shape of the wind profile), environments with larger near-ground streamwise vorticity yield more intense low-level mesocyclones. In such cases, additional horizontal vorticity from the forward flank is superfluous to establishing the footprint of dynamic lifting in the low-levels of a supercell.  

These results may run counter to the growing interest in SVCs \citep[i.e.,][]{orf2017evolution}. The tornadic supercell simulated by \citet{orf2017evolution}, and expanded upon in \citet{finley2023svc}, possessed a base-state environment with an incredible amount of near-ground streamwise vorticity (SRH500 $>$ 300 m$^{2}$ s$^{-2}$), and it is not yet known how much the storm-generated vorticity present in the SVC contributed to the low-level mesocyclone (compared to the stretching of ambient, environmental vorticity) or whether the absence of an SVC would have hindered tornadogenesis. It seems possible the development of an SVC could be the manifestation of air accelerating towards an intensifying low-level mesocyclone, commiserate with near-ground mass continuity, and the associated horizontal stretching yielding enhanced streamwise vorticity. In a similarly high-resolution tornadic supercell simulation in a different environment, \citet{finley2023svc} also found that a tornado formed prior to forward flank cold pool development, and the SVC seemed to have little influence on the tornadogenesis process, with a mature SVC not developing until well after the tornadogenesis had occurred. Even if SVCs are ubiquitous in surface-based supercells \citep[regardless of tornado outcome and/or intensity, e.g.,][]{markowski2020intrinsic}, those augmented parcels need to participate in the low-level mesocyclone. This seems not to be the primary pathway for the development of low-level mesocyclones in our simulations or the recent studies by \citet{murdzek2020svc,murdzek2020processes}. The orientation of the wind profile, in relation to the updraft's inflow low and baroclinic vorticity produced in the forward flank, might also determine whether these air parcels contribute more or less to the low-level mesocyclone \citep{schueth2021svc}. Future research could focus on the quantitative contribution of the SVC to the low-level mesocyclone across a range of storms and environments and whether a storm-generated SVC is an instigator of tornadogenesis (or more related to tornado intensification and/or maintenance) or whether the SVC is primarily a by-product of an environmentally derived low-level mesocyclone.  

As with any modeling study, the scope of our results could depend on numerous, untested factors, including (but not limited to) a narrow range of thermodynamic environments, the magnitude of low-level cooling attributable to the microphysics scheme, and the possibility that a single simulation in each wind profile is not representative of what a larger ensemble would produce. To this end, we have presented results from simulations of both \citet{wade2021dynamics} and \citet{peters2023disentangling}, which used different environments, microphysics schemes, treatments of the lower boundary condition, and sub-grid turbulence closures. Even within a set of SOM-generated hodographs designed to find distinct patterns in the 0-500 m wind profile, there appear to be factors at play that exert some control over the tornado outcome beyond simply the ingestion of large near-ground SRH. Two such examples include: 1) the 0\_0 supercell producing a tornado despite being in an environment with SRH500 well below the tornadic value from \citet{coffer2019srh500}; 2) the 0\_2 supercell being violently tornadic despite an environmental SRH500 value in the lower inner-quartile range for significantly tornadic supercells from \citet[their Fig. 2]{coffer2019srh500}, in addition to a larger component of crosswise vorticity in the lower troposphere \citep[an environmental trait known to be unfavorable for steady low-level mesocyclone;][]{coffer2018tipping}.”

Some of the outstanding questions related to this study that remain to be addressed include: Why do the weaker supercells (such as-in 1\_0 herein and in \citet{peters2023disentangling}, see Supplemental Fig. 7e,f) have more forward flank baroclinic contributions? Are these storms weaker because the baroclinic contribution is high or is this a symptom of something else wrong in the storm? Additionally, is our finding that the dominant contributor to vertical vorticity within the low-level mesocyclone is from the environmental horizontal vorticity due to the lack of baroclinity in the forward flank of these storms (and in observed tornadic storms) or is this more to due with the trajectory shapes themselves? Even if there were larger buoyancy gradients and more baroclinic generation of streamwise vorticity, would the additional negative buoyancy still favor inflow parcels from the ambient environment? In the future, consideration should be given to whether the storm generated vorticity contributes more within storms with larger buoyancy gradients within the forward flank than presented herein.    

In conclusion, we posit a potential connection between the environmental streamwise vorticity and the potential role of the storm-generated streamwise vorticity. In order for rain-cooled air (such as the SVC) to be re-ingested by the storm’s low-level updraft in the first place, substantial dynamic lifting must be present to force these outflow parcels with negative buoyancy upwards to their levels of free convection. There appears to be a threshold of streamwise vorticity, above which, the dynamical response to streamwise horizontal vorticity increases dramatically \citep{coffer2018tipping,goldacker2021updraft}. We hypothesize that the ambient environmental streamwise vorticity establishes a ``floor'' of low-level mesocyclone intensity (and tornadogenesis potential), which could then be supplemented by within-storm baroclinic generation of streamwise vorticity. In high streamwise vorticity environments, the probability of tornadogenesis is argued to be high because very little (or no) augmentation from the storm is required in order to instigate the dynamical updraft response that favors tornadogenesis. %; this would make storms rather insensitive to within-storm processes. 
On the other hand, in low streamwise vorticity environments, the probability of tornadogenesis is argued to be lower because much more augmentation from the storm is required \citep[or from external sources such as storm mergers, e.g.,][]{fischer2022mergers}. 
%The probability is non-zero of course and the potential influx of storm-generated SRH may in part explain how supercells in seemingly sub-optimal environments occasionally produce tornadoes (perhaps such as-in the 0\_0 supercell). % (as well as large degree of overlap between the environments supporting tornadic and nontornadic supercells). %; this would make storms comparatively sensitive to within-storm processes.
The necessity of within-storm supplementation adds to the possible failure points, as there must be favorable storm interactions and/or sufficient dynamic lifting to overcome the negative buoyancy associated with the baroclinic vorticity generation. We plan to further explore supercell simulations in both marginal environments and environments with a wider range of thermodynamic profiles to test this hypothesis. Regardless, we believe the results of this study will fundamentally clarify previous ambiguity within the literature on low-level mesocyclones in supercells. 

%%%%%%%%%%%%%%%%%%%%%%%%%%%%%%%%%%%%%%%%%%%%%%%%%%%%%%%%%%%%%%%%%%%%%
% ACKNOWLEDGMENTS
%%%%%%%%%%%%%%%%%%%%%%%%%%%%%%%%%%%%%%%%%%%%%%%%%%%%%%%%%%%%%%%%%%%%%
\acknowledgments
We thank the anonymous peer reviewers for their thought-provoking feedback on this manuscript, which challenged some of our initial assumptions and ultimately led to an improved, more complete, end product. B. Coffer's and M. Parker's efforts were supported by the National Science Foundation (NSF) on grant AGS-2130936. J. Peters's efforts were supported by NSF grants AGS-1928666, AGS-1841674, and the Department of Energy Atmospheric System Research (DOE ASR) grant DE-SC0000246356. A. Wade was supported by National Oceanic and Atmospheric Administration(NOAA)–University of Oklahoma Cooperative Agreement NA16OAR4320115. Dr. George Bryan is gratefully acknowledged for providing and supporting the CM1 model. We appreciate the assistance of Dr. Sonia Lasher-Trapp with code for the heat flux initialization in CM1. Dr. Johannes Dahl is acknowledged for general guidance of implementing stencils. Nicholas Goldacker supplied the hodographs from the SOM (doi:10.5061/dryad.f1vhhmh1fthe9), and Adam Werkerma developed the mesocyclone tracking algorithm. The base code for backward trajectories was based on Dr. Tom Gowan’s GitHub repository (https://tomgowan.github.io/). The authors also thank John Hart, Ryan Jewell, Bryan Smith, and Rich Thompson of the NOAA/NWS Storm Prediction Center for their work on the underlying software used to generate the SFCOA proximity sounding database, tireless effort in assigning convective modes to each report, and their generosity for sharing the data with the community. Drs. George Bryan, Johannes Dahl, Mike Coniglio, Lou Wicker, and Patrick Skinner, as well as Kevin Gray and Alex Schueth provided useful analysis ideas and feedback on a preliminary version of this work. Many figures and analyses were created using open-sourced python packages such as Numpy (Harris et al. 2020), Matplotlib (Hunter 2017), Pandas (Wes McKinney 2010), Xarray (Hoyer and Hamman 2016), Metpy \citep{may2017metpy}, and Jupyter (Kluyver et al. 2016), as well as the Grid Analysis and Display System (GrADS). Finally, we would like to acknowledge high-performance computing support from Cheyenne (doi:10.5065/D6RX99HX) provided by NCAR's Computational and Information Systems Laboratory, sponsored by NSF.

%%%%%%%%%%%%%%%%%%%%%%%%%%%%%%%%%%%%%%%%%%%%%%%%%%%%%%%%%%%%%%%%%%%%%
% DATA AVAILABILITY STATEMENT
%%%%%%%%%%%%%%%%%%%%%%%%%%%%%%%%%%%%%%%%%%%%%%%%%%%%%%%%%%%%%%%%%%%%%
\datastatement
The CM1 model code is available from https://www2.mmm.ucar.edu/people/bryan/cm1/. All other namelists, input files, and post-processing scripts are available from the author by request.

%%%%%%%%%%%%%%%%%%%%%%%%%%%%%%%%%%%%%%%%%%%%%%%%%%%%%%%%%%%%%%%%%%%%%
% REFERENCES
%%%%%%%%%%%%%%%%%%%%%%%%%%%%%%%%%%%%%%%%%%%%%%%%%%%%%%%%%%%%%%%%%%%%%
% Make your BibTeX bibliography by using these commands:
\bibliographystyle{ametsocV6}
\bibliography{references}

\ifarXiv
    

\section*{Supplementary material}

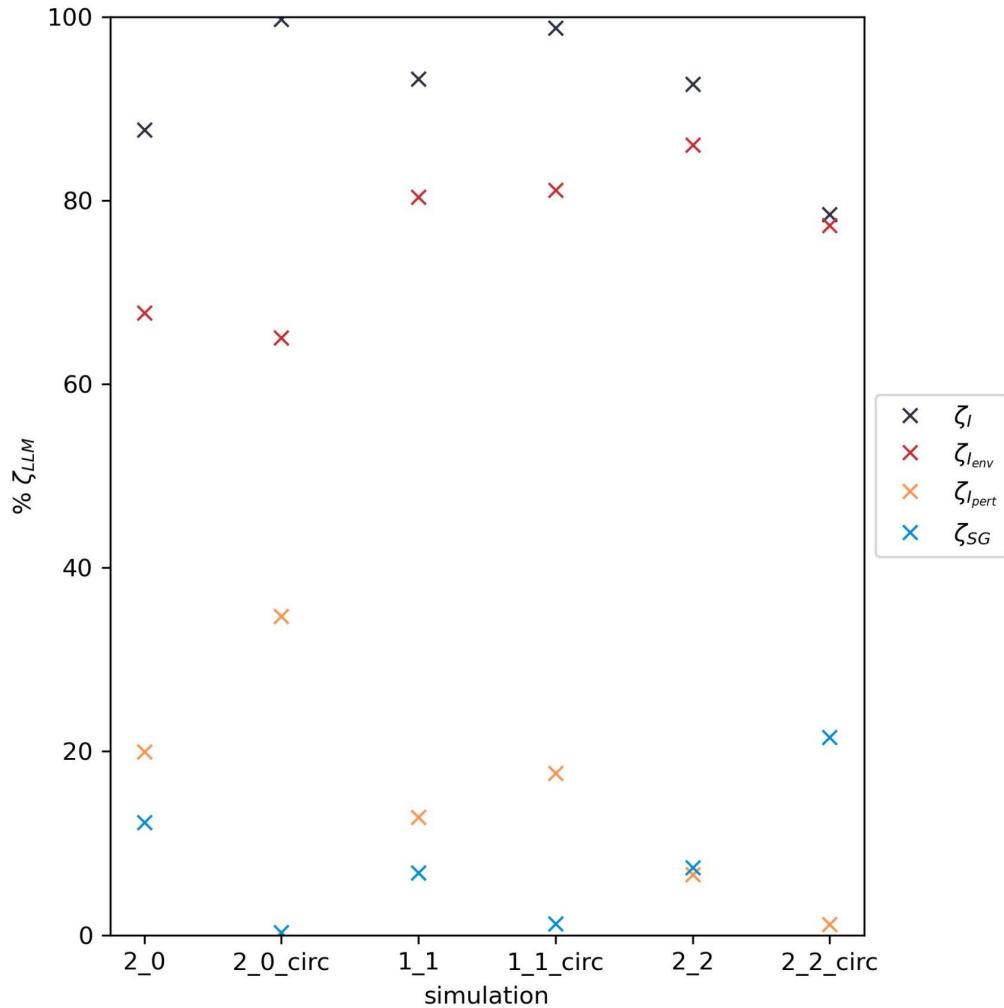

Supplemental Figure 5: Percentage of $\zeta_{LLM}$ attributable to $\zeta_I$, $\zeta_{Ienv}$, $\zeta_{Ipert}$, and $\zeta_{SG}$ for the circulation based alternative low-level mesocyclone definitions ("_circ") for three supercell simulations, representing one of each class of tornado outcomes: nontornadic (2_0), tornadic (1_1), and violently tornadic (2_2). These supercells were chosen due to their well behaved manner as right-moving storms but also because the base-state SRH500 values were close to the median climatological values for nontornadic, weakly tornadic, and significantly tornadic right-moving supercells from Coffer et al. (2019, see their Table 2).. The `x' marks the median percentage value from all the low-level mesocyclone trajectories for each simulation. Similar to Fig. 13 except trajectories were filtered for the 90th percentile of circulation at 1 km AGL (with no vertical vorticity or velocity thresholds requirements).



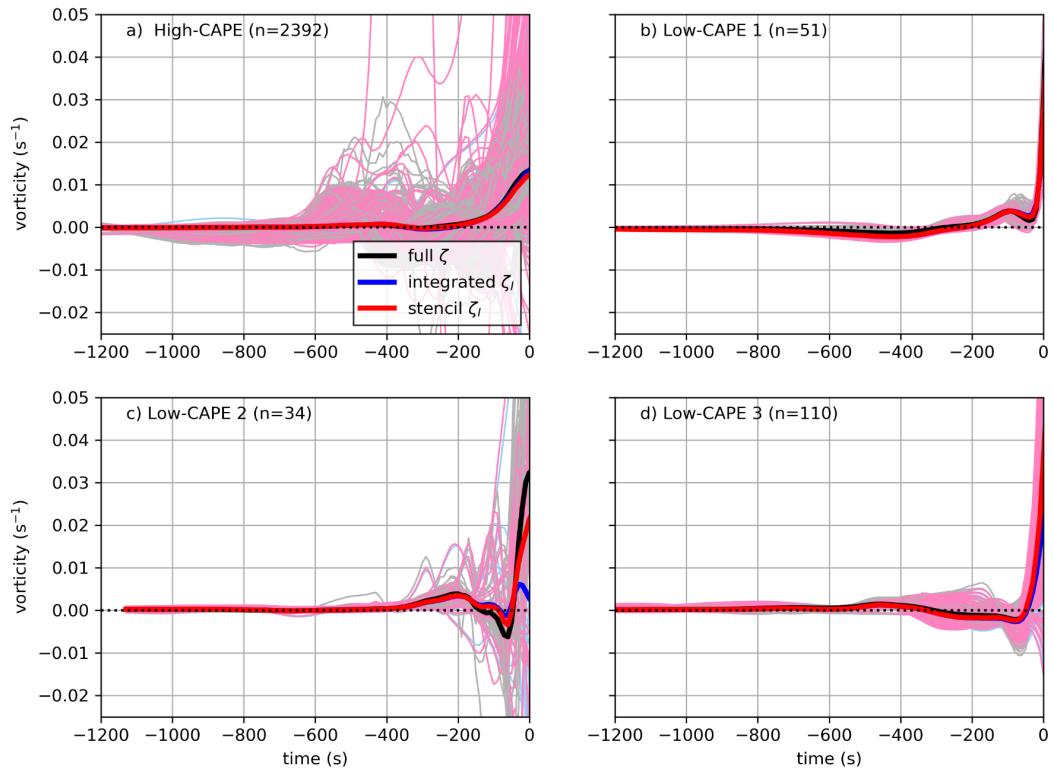

Supplemental Figure 6: Time series of vertical vorticity for all low-level mesocyclone parcels (fine lines) and parcel group means (bold) for the full vertical vorticity (black) and the initial vertical vorticity ($\zeta_I$) of the parcel determined via integrating the vorticity equation (blue) and via the material stencil method (red). Dashed lines are the mean base-state vertical vorticity. Each panel represents a different supercell simulation from Wade and Parker (2021). Adapted from Wade (2020).



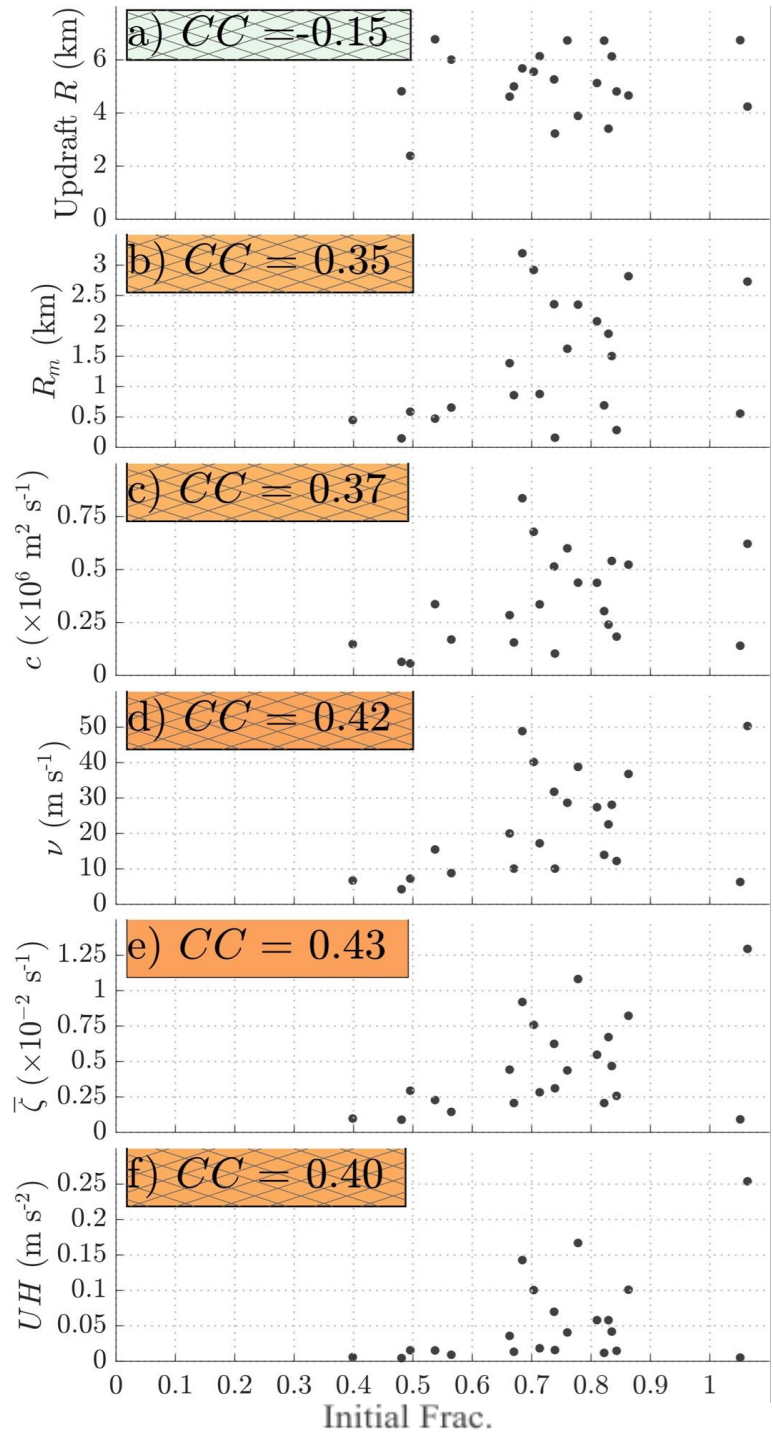

Supplemental Figure 7: Scatter plot of the initial fraction of the low-level mesocyclone ($\zeta_l$ / $\zeta_{LLM}$) on the abscissa compared to updraft radius (a), mesocyclone radius (b), net mesocyclone circulation (c), net mesocyclone rotational velocity (d), average mesocyclone vertical vorticity (e), and average mesocyclone updraft helicity (e) along the ordinate from the simulations by Peters et al. (2023). Spearman correlation coefficients $CC$ are shown in panel labels, with the panel label color reflecting the relative magnitude of $CC$ (reds are positive, blues are negative). Hatched $CC$ are not statistically significant, based on a student's t-test.



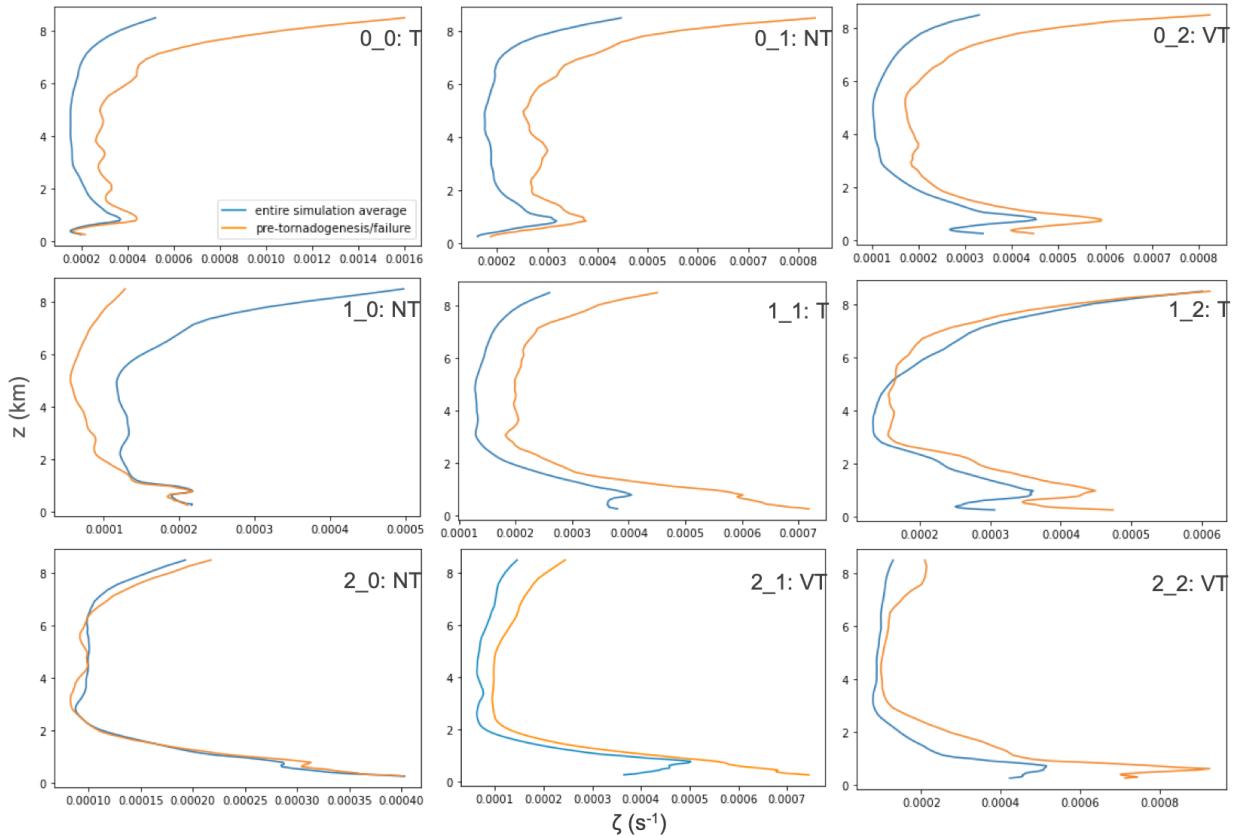

Supplemental Figure 8: Domain-wide 90th percentile of vertical vorticity as a function of height for the supercell simulations presented herein, during both the 1-3 hour time range (i.e., the entire duration of the simulation at the downscaled grid-spacing of 80 m), excluding tornadic periods, and the key time periods of tornado-genesis/failure defined in Section 2c-2. Each panel is labeled nontornadic (NT), tornadic (T), or violently tornadic (VT).


\fi

\end{document}